\documentclass{ieeeaccess}
\usepackage{cite}
\usepackage{amsmath,amssymb,amsfonts}
\usepackage{algorithmic}
\usepackage{graphicx}
\usepackage{textcomp}
\usepackage{soul}
\usepackage[hidelinks]{hyperref}
\usepackage{graphicx}
\usepackage{subfig}
\usepackage{lipsum}
\usepackage{multirow}
\usepackage[table,xcdraw]{xcolor}
\usepackage{lipsum}
\usepackage{caption}
\usepackage[linesnumbered,ruled,vlined]{algorithm2e}
\SetKwInput{KwInput}{Input}                
\SetKwInput{KwOutput}{Output}              

\def\doubleunderline#1{\underline{\underline{#1}}}

\DeclareMathOperator*{\argmax}{argmax}  
\DeclareMathOperator*{\argmin}{argmin}  
\newcommand{\MKH}[1]{\textcolor{black}{#1}}
\usepackage{enumerate}

\newsavebox\MBox
\newcommand\Cline[2][blue]{{\sbox\MBox{$#2$}%
  \rlap{\usebox\MBox}\color{#1}\rule[-3.9\dp\MBox]{\wd\MBox}{0.95pt}}}

\def\BibTeX{{\rm B\kern-.05em{\sc i\kern-.025em b}\kern-.08em
    T\kern-.1667em\lower.7ex\hbox{E}\kern-.125emX}}
\begin{document}
\history{Date of publication xxxx 00, 0000, date of current version xxxx 00, 0000.}
\doi{10.1109/ACCESS.2017.DOI}

\title{Multi-class probabilistic atlas-based whole heart segmentation method in cardiac CT and MRI}
\author{\uppercase{Tarun Kanti Ghosh}\authorrefmark{1},
\uppercase{Md. Kamrul Hasan}\authorrefmark{2}, 
\uppercase{Shidhartho Roy}\authorrefmark{2},
\uppercase{Md. Ashraful Alam}\authorrefmark{2},
\uppercase{Eklas Hossain}\authorrefmark{3} \IEEEmembership{Senior Member, IEEE}, 
\uppercase{Mohiuddin Ahmad}\authorrefmark{2} \IEEEmembership{Member, IEEE}
}
\address[1]{Department of Biomedical Engineering, Khulna University of Engineering \& Technology, Khulna-9203, Bangladesh}

\address[2]{Department of Electrical and Electronic Engineering, Khulna University of Engineering \& Technology, Khulna-9203, Bangladesh}

\address[3]{Department of Electrical Engineering \& Renewable Energy, Oregon Renewable Energy Center (OREC), Oregon Institute of Technology, OR 97601, USA}


\markboth
{Ghosh \headeretal: Multi-class probabilistic atlas-based whole heart segmentation method in cardiac CT and MRI}
{Ghosh \headeretal: Multi-class probabilistic atlas-based whole heart segmentation method in cardiac CT and MRI}

\corresp{Corresponding author: Md. Kamrul Hasan (e-mail: m.k.hasan@eee.kuet.ac.bd).}

\begin{abstract}
Accurate and robust whole heart substructure segmentation is crucial in developing clinical applications, such as computer-aided diagnosis and computer-aided surgery.
However, segmentation of different heart substructures is challenging because of inadequate edge or boundary information, the complexity of the background and texture, and the diversity in different substructures' sizes and shapes.
This article proposes a framework for multi-class whole heart segmentation employing non-rigid registration-based probabilistic atlas incorporating the Bayesian framework.
We also propose a non-rigid registration pipeline utilizing a multi-resolution strategy for obtaining the highest attainable mutual information between the moving and fixed images.    
We further incorporate non-rigid registration into the expectation-maximization algorithm and implement different deep convolutional neural network-based encoder-decoder networks for ablation studies.
All the extensive experiments are conducted utilizing the publicly available dataset for the whole heart segmentation containing $20$ MRI and $20$ CT cardiac images.
The proposed approach exhibits an encouraging achievement, yielding a mean volume overlapping error of $14.5\,\%$ for CT scans exceeding the state-of-the-art results by a margin of $1.3\,\%$ in terms of the same metric.
As the proposed approach provides better-results to delineate the different substructures of the heart, it can be a medical diagnostic aiding tool for helping experts with quicker and more accurate results.

\end{abstract}


\titlepgskip=-15pt

\maketitle

\section{Introduction}
\label{introduction}
In this section, we first specify the challenges and motivations of this article in subsection~\ref{Problem_presentation}. Secondly, we review several recent literature in subsection~\ref{Recent_methods} for the particularized difficulties in subsection~\ref{Problem_presentation}. Finally, in subsection~\ref{Our_contribution}, we summarize our contributions in this article.

\subsection{Problem presentation}
\label{Problem_presentation}
Medical imaging and computing technologies have revolutionized modern medicine and healthcare, which are increasingly crucial for the treatments and diagnosis of different Cardiovascular Diseases (CVDs) \cite{zhuang2019evaluation, mendis2011global}.
Currently, different non-invasive cardiac imaging assessments, such as Magnetic Resonance Imaging (MRI), Computed Tomography (CT), Ultrasound (US), Positron Emission Tomography (PET), and Single Photon Emission Computed Tomography (SPECT), are being used widely for clinical and diagnostic applications in cardiology \cite{kang2012heart, liu2019automatic}. 
Among all of those imaging modalities, the MRI and CT scans have essential functions in the non-invasive evaluation of CVDs through extensive experimentation, and clinical applications \cite{zhuang2019evaluation, mortazi2017multi}. 
The CT images are more commonly employed than MRI images due to their high-speed retrieval and more affordable expense, whereas the MRI images have no ionizing radioactivity and outstanding soft-tissue contrast.
However, most of the current clinically convenient image examination approaches either attunes for the MRI or CT images alone \cite{zhuang2019evaluation, mortazi2017multi}.

Delineating different vital volumetric substructures from the whole volumetric medical images are of great significance for clinical practice for quantifying the morphological and pathological changes \cite{zhuang2019evaluation}.
Precise segmentation aids the subsequent quantitative evaluation of the Volume of Interests (VOIs). It also avails specific diagnosis, forecast of prognosis, computer-aided diagnosis, radiation therapy, computer-aided surgery, intra-operative guidance, and surgical planning \cite{zhuang2019evaluation, dou20173d}.
For instance, liver segmentation from $3D$ abdominal CT scans is a crucial prerequisite for tumor resection,  computer-assisted live donor transplantation, and minimally invasive surgery interventions \cite{heimann2009comparison, radtke2007computer, meinzer2002computerized}. Besides, for the treatment of CVDs, including radio-frequency ablation and surgical preparation of crucial congenital heart diseases, volumetric cardiac MR image segmentation is necessary \cite{peters2007automatic, pace2015interactive, atehortua2016automatic}.

\begin{figure*}[!ht]
  \centering
  \subfloat{\includegraphics[width=17cm, height=5.1cm]{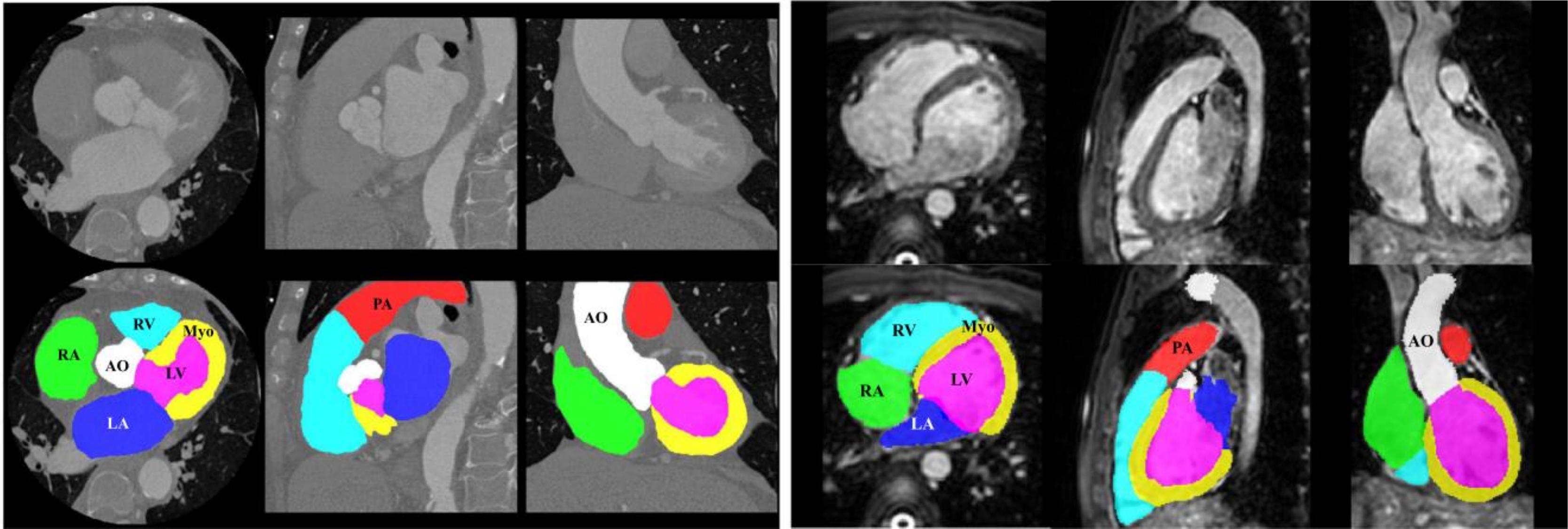}} 
  \caption{Illustration of complexities for achieving robust WHS in the employed cardiac MM-WHS-2017 dataset with corresponding ground-truth, borrowed from \cite{zhuang2019evaluation}, where the left three columns display the three orthogonal views of a cardiac CT image and its corresponding WHS result, and the right three columns exhibit example cardiac MRI data and the WHS result. The LV, RV, LA, RA, Myo, AO, PA, and background in the first columns have a different shape, texture, and spatial location than the other two CT columns (second and third). Similar complexities are notified for the MRI scans (last three columns).}
  \label{fig:dataset}
\end{figure*}

Currently, the Whole Heart Segmentation (WHS) is an imperative preliminary action for a wide range of clinical treatments. For example, the pathology localization and accurate ventricular dimensions \cite{zhao2019accurate, de2020fast}, which aims to delineate seven different heart substructures, as outlined in Table~\ref{tab:whs_sub}, from the whole cardiac images (see in Fig.~\ref{fig:dataset}). 
\begin{table*}[!ht]
\centering
\caption{Details of different heart substructures to be segmented from the whole chest images and their anatomical locations.}
\label{tab:whs_sub}
\begin{tabular}{lll}
\hline
\rowcolor[HTML]{C0C0C0} 
Substructures    & Acronym     & Anatomical position and functions                                                                   \\ \hline  

Left Ventricular cavity          & LV               & The bottom left portion of the heart below the left atrium for pumping oxygenated blood to all tissues                            \\ 

Right Ventricular cavity         & RV               & The bottom right portion of the heart below the right atrium for pumping oxygen-depleted blood to the lungs\\

Left Atrial cavity               & LA               & 
The upper left portion of the heart to receive oxygenated blood from the four pulmonary veins

\\ 
Right Atrial cavity              & RA               & The upper right portion of the heart for returning deoxygenated blood from the body to the RV             \\

Myocardium of LV & Myo              & The myocardium is the muscular middle layer of the wall of the heart for pumping blood around the body

\\ 
Ascending Aorta                  & AO               &   The ascending aorta is connected to the heart's LV to allow the flow of blood from the heart into the aorta \\ 

Pulmonary Artery                 & PA               &   The pulmonary artery begins at the base of the heart’s RV to deliver oxygen-depleted blood to each similar lung                               \\ \hline
\end{tabular}
\end{table*}
The precise heart quantification requires the subtle segregation of different heart substructures. 
For instance, the ejection portion and the myocardial mass are estimated from the segmented ventricular and the myocardial results, respectively, which are critical pointers of cardiac disorder detection.
The whole heart's manual delineation is labor-intensive and tedious, necessitating almost eight hours for an individual subject \cite{zhuang2016multi}.  
Consequently, the designing of computer-aided methods to investigate medical images automatically are highly demanding. 
However, an automated segmenting of a whole heart is also challenging due to the cardiac anatomical shape variations and the indefinite borders among different heart substructures \cite{zhuang2019evaluation}, as depicted in Fig.~\ref{fig:dataset}. 
Achieving entirely computerized WHS is arduous due to the following hurdles:

\begin{itemize}
    \item Wide shape, structure, and boundary variations of the cardiac anatomy and their subsequent cardiac imaging
    
    \item Indefinite boundaries, with inadequate edge information, between different heart substructures in cardiac images
    
    \item Low cardiac image quality and seldom they blend with other visually alike substructures (see in Fig.~\ref{fig:dataset}) 
\end{itemize}

\subsection{Recent methods}
\label{Recent_methods}
Convolution Neural Network (CNN)-based Deep Learning (DL) strategies are the most commonly employed medical image segmentation method in different application fields and modalities \cite{hasan2020drnet, hasan2020dsnet, vigneault2018omega, xu2018cfun, ding2020cab, han2020cascaded}. 
Table~\ref{tab:different_Methods} bestows various DL-based and other techniques for the WHS of different substructures, including their corresponding utilized datasets and identical outcomes.
\begin{table*}[!ht]
\caption{Several published literature for WHS with their employed datasets and achievements conferring varying metrics, such as mDSC, mVOE, and mSn, respectively, for mean dice similarity coefficient, mean volume overlapping error, and mean sensitivity.}
\centering
\begin{tabular}{p{10cm}clccc}
\hline
\rowcolor[HTML]{C0C0C0} 
\cellcolor[HTML]{C0C0C0}                                    & \cellcolor[HTML]{C0C0C0}                       & \cellcolor[HTML]{C0C0C0}                           & \multicolumn{3}{c}{\cellcolor[HTML]{C0C0C0}Metrics} \\ \cline{4-6} 
\rowcolor[HTML]{C0C0C0} 
\multirow{-2}{*}{\cellcolor[HTML]{C0C0C0}Different methods} & \multirow{-2}{*}{\cellcolor[HTML]{C0C0C0}Year} & \multirow{-2}{*}{\cellcolor[HTML]{C0C0C0}Datasets} & mDSC             & mVOE            & mSn            \\ \hline

\multirow{2}{10cm}{A multi-modality atlas with a distinct label merging method based on a multi-scale patch approach and a unique global atlas ranking system  \cite{zhuang2016multi}}

  & \multirow{2}{*}{$2016$} &      \multirow{2}{*}{MM-WHS-2017}                                                                                       &   \multirow{2}{*}{$0.899$}  &   \multirow{2}{*}{$0.183$} &   \multirow{2}{*}{$-$ }              \\ \hline

  \multirow{2}{10cm}{An automated approach using dilated convolutional neural networks for aggregating features at various scales through convolutional layers with too fewer parameters \cite{wolterink2016dilated}}

  & \multirow{2}{*}{$2016$} &      \multirow{2}{*}{HVSMR-2016}                                                                                       &   \multirow{2}{*}{$0.865$}  &   \multirow{2}{*}{$0.238$} &   \multirow{2}{*}{$-$ }              \\ \hline

\multirow{2}{10cm}{A multi-planar deep CNNs with an adaptive merging procedure utilizing corresponding information from the separate planes of the 3D scans for enhanced delineations\cite{mortazi2017multi}}  
   
  & \multirow{2}{*}{$2017$} &      \multirow{2}{*}{MM-WHS-2017}                                                                                       &   \multirow{2}{*}{$0.851$}  &   \multirow{2}{*}{$0.259$} &   \multirow{2}{*}{ $0.831$}              \\ \hline

\multirow{2}{10cm}{An atlas-based segmentation approach with a two-stage registration pipeline, where the majority voting and STEPS algorithms were used for the merging of
the labels \cite{zuluaga2016strengths}}  

  & \multirow{2}{*}{$2017$} &      \multirow{2}{*}{HVSMR-2016}                                                                                       &   \multirow{2}{*}{$0.900$}  &   \multirow{2}{*}{$0.182$} &   \multirow{2}{*}{$-$ }              \\ \hline

\multirow{3}{10cm}{A method having three crucial actions: heart identification from landmark detection, heart isolation using mathematical shape model, and segmentation using learning-based voxel classification and local phase analysis \cite{wang2016automatic}}  
   
  & \multirow{3}{*}{$2018$} &      \multirow{3}{*}{\MKH{HVSMR-2016}}                                                                                       &   \multirow{3}{*}{$0.760$}  &   \multirow{3}{*}{$0.387$} &   \multirow{3}{*}{$-$ }              \\ \\ \hline

\multirow{2}{10cm}{A deeply-supervised fractal network, where the multi-paths with different receptive fields were established in a self-similar fractal system to obtain the hierarchical features \cite{yu20163d}}  
   
  & \multirow{2}{*}{$2018$} &      \multirow{2}{*}{\MKH{HVSMR-2016}}                                                                                       &   \multirow{2}{*}{$0.760$}  &   \multirow{2}{*}{$0.387$} &   \multirow{2}{*}{$-$ }              \\ \hline

\multirow{2}{10cm}{Two-stage UNet framework, where the first stage detected the VOIs and the second stage accurately segmented the heart substructures \cite{liu2019automatic}}  
   
  & \multirow{2}{*}{$2019$} &      \multirow{2}{*}{MM-WHS-2017}                                                                                       &   \multirow{2}{*}{$0.793$}  &   \multirow{2}{*}{$0.342$} &   \multirow{2}{*}{$-$ }              \\ \hline

\multirow{2}{10cm}{A semi-supervised approach, where the student model acquires from labeled target data and also searches unlabeled target data and labeled data by two teacher models \cite{li2020dual}}  
   
  & \multirow{2}{*}{$2019$} &      \multirow{2}{*}{\MKH{MM-WHS-2017}}                                                                                        &   \multirow{2}{*}{$0.860$}  &   \multirow{2}{*}{$0.245$} &   \multirow{2}{*}{$-$ }              \\ \hline

\multirow{2}{10cm}{Category attention boosting module connecting the deep network estimation graph with the boosting approach \cite{ding2020cab}}  

                       & \multirow{2}{*}{$2019$} &     \multirow{2}{*}{HVSMR-2016}   &   \multirow{2}{*}{$0.894$}   &  \multirow{2}{*}{$0.191$}  &  \multirow{2}{*}{$-$}  \\ \hline
                       
\multirow{2}{10cm}{A CNN-based architecture, which includes principal component analysis as a supplementary data enlargement routine \cite{habijan2019whole}}  
   
  & \multirow{2}{*}{$2019$} &      \multirow{2}{*}{MM-WHS-2017}                                                                                       &   \multirow{2}{*}{$0.890$}  &   \multirow{2}{*}{$0.198$} &   \multirow{2}{*}{$-$ }              \\ \hline

\multirow{3}{10cm}{A deep heterogeneous feature gathering network (HFANet) to wholly employ corresponding knowledge from various views of 3D cardiac data \cite{zheng2019hfa}}  & $2019$ &       HVSMR-2016    &   $0.942$ &  $0.110$  &  $-$   \\

                       & $2019$ &   MM-WHS-2017    &   $0.909$   &  $0.167$  &  $-$ \\
                       & $2019$ &     AAPM-2017   &   $0.883$   &  $0.209$  &  $-$\\ \hline

\multirow{2}{10cm}{An adversarial training strategy for training the networks using the UNet as a generator in adversarial network \cite{dong2019automatic}}  
   
  & \multirow{2}{*}{$2020$} &      \multirow{2}{*}{HVMSR-2016}                                                                                       &   \multirow{2}{*}{$0.870$}  &   \multirow{2}{*}{$0.230$} &   \multirow{2}{*}{$0.890$ }              \\ \hline

\multirow{2}{10cm}{Cascading of two volumetric FCNs, where the first network aimed at locating the cardiac area, during the second segmented different cardiac and great vessel substructures \cite{han2020cascaded}}  
   
  & \multirow{2}{*}{$2020$} &      \multirow{2}{*}{HVSMR-2016}                                                                                       &   \multirow{2}{*}{$0.942$}  &   \multirow{2}{*}{$0.109$} &   \multirow{2}{*}{$0.943$ }              \\ \hline

\multirow{2}{10cm}{ A pipeline of two FCNs, where the first network localizes the bounding box's center encompassing all heart substructures, while the second network segments them  \cite{payer2017multi}}  
   
  & \multirow{2}{*}{$2020$} &      \multirow{2}{*}{MM-WHS-2017}                                                                                       &   \multirow{2}{*}{$0.908$}  &   \multirow{2}{*}{$0.168$} &   \multirow{2}{*}{$-$ }              \\ \hline
  
\end{tabular}
\label{tab:different_Methods}
\end{table*}
UNet, initially proposed in \cite{ronneberger2015u}, is mostly applied CNN structure for the WHS \cite{vigneault2018omega, dong2019automatic, xu2018cfun, wang2017automatic, tong20173d, ye2019multi, liu2019automatic, habijan2019whole, ding2020cab}, which can be summarized as follows: 

The authors in \cite{habijan2019whole} suggested a framework consisting of two 3D-UNets, where the first network was employed to localize the bounding box encompassing the heart, and the second network was used for the fine segmentation of different substructures. They also employed principal component analysis-based image augmentation for enhancing the WHS results.
An UNet-based Omega-Net was introduced in \cite{vigneault2018omega} consisting of a set of UNet for fine-grained WHS. By explicitly finding the VOIs and turning the input image to a standard orientation, this method adequately accommodated the loss of segmentation precision produced by distinguishing between training and testing images. 
Faster R-CNN (FRCNN) and UNet were combined in \cite{xu2018cfun} naming as a CFUN. With the FRCNN's precise localization ability and UNet's strong segmentation capability, CFUN needed solely one-step detection and segmentation deduction to receive the VOIs. The authors also used a novel loss function based on border information to accelerate the training and enhance the WHS results.
The author in \cite{wang2017automatic} applied an approach combining with shape context knowledge encoded in volumetric shape models consisting of three primary steps: explorer segmentation with orthogonal 2D-UNet, shape context estimation, and refining segmentation with UNet and shapes context. The additional shape information, also called shape context, was applied for implementing explicit 3D shape knowledge to CNN.
A multi-depth fused 3D-UNet was applied in \cite{ye2019multi} to the initial network for more reliable extracting context information. The authors also introduced a hybrid loss incorporating focal loss into the dice function to mark the volume size asymmetry among various ventricular substructures. 
The authors in \cite{tong20173d} developed a pipeline with three quintessential steps: employment of 3D-UNet for coarse detection of VOIs to alleviate surrounding tissues' impact, artificially augmentation of the training dataset by extracting different VOIs, and a refined 3D-UNet for segmentation refinement using the augmented training dataset. 
The authors in \cite{ding2020cab} joined the attention tool into the gradient expanding process for enhancing the coarse segmentation information with less computation expense. Moreover, they introduced the Category Attention Boosting (CAB) module into the 3D-UNet network and constructed a new multi-scale boosting model, CAB-UNet, extending the network's gradient flow and making sole usage of the low-resolution feature information.

However, the segmentation strategies relying on the UNet-based structure suffer from redundant fusion tactics, imposing concatenation only at the identical scale feature maps of the encoder and decoder \cite{hasan2020drnet, hasan2020dsnet, zhou2019unet++}.
There are some other DL- and Atlas-based techniques for the WHS \cite{mo2018deep, mortazi2017multi, payer2017multi, zheng2019hfa, li2020dual, yu20163d, han2020cascaded, zhuang2016multi}, which are also reviewed and summarized as follows:

A deep poincaré map encapsulating prior knowledge with a dynamical system was applied in \cite{mo2018deep}. A CNN-based approach had then navigated an ambassador over the cardiac MRI image, moving to approach a route that outlined the target substructures. 
Different CT or MRI plans, such as coronal, sagittal, and axial, are explored for an adaptive fusion in \cite{mortazi2017multi} with a multi-planar deep CNN-based method. The authors independently built and trained three CNNs, with the same constructive arrangement, for those three planes and finally, combined them to obtain the segmented VOIs.
A pipeline having two Fully Convolutional Networks (FCNs) was implemented in \cite{payer2017multi}, where the first CNN localized the bounding box's center around the different substructures and the succeeding second CNN concentrated on these regions for segmenting the organs. 
The authors in \cite{zheng2019hfa} developed a deep Heterogeneous Feature Aggregation Network (HFANet) for entirely exploiting corresponding information from 3D cardiac data. They utilized asymmetrical 3D kernels and pooling for obtaining heterogeneous features in identical encoding routes. Therefore, distinguishable features were extracted from a specific view, and necessary contextual information was kept. Then, they employed a content-aware multi-planar concatenation for aggregating meaningful features to boost the WHS performance. Further, to overcome the model size, they also devised a new DenseVoxNet model by sparsifying skip connections trained in an end-to-end manner.
A Dual-Teacher strategy was suggested in \cite{li2020dual}, where the student model acquired precisely the knowledge of unlabeled target data from intra-domain teachers by fostering prediction texture and the shape priors embedded in labeled source data from inter-domain teachers via information distillation. The authors also examined the utility of concurrently leveraging unlabeled data and well-known cross-modality data for the segmentation. 
A 3D fractal network for effective computerized segmentation technique was introduced in \cite{yu20163d}. The designed network took full convolutional construction to implement effective, well-defined, and volume-to-volume prognostication. Prominently, by recursively employing a single augmentation rule, the authors assembled the network in a novel self-similar fractal system and consequently promoted it in consolidating hierarchical evidence for precise segmentation. They also employed a deep supervision tool to mitigate the vanishing gradients obstacle and increase our network's training effectiveness on inadequate medical image datasets.  
The authors in \cite{zhuang2016multi} offered a multi-scale patch for hierarchical local atlas ranking.  Their segmentation approach exercised multi-modality atlases from MRI and CT and embraced a new label merging method based on the recommended multi-scale patch policy and a new global atlas ranking scheme. Both the local and global atlas ranking actions used the information-theoretic criteria to estimate the association between the target image and the atlases from versatile modalities.

\subsection{Our contribution}
\label{Our_contribution}
While many approaches have already been developed and implemented for the WHS, there is still room for performance improvement. This article proposes a statistical WHS method that joins the prior anatomical information described by probabilistic atlas into the Bayesian inference for delineating seven different heart substructures (see in subsection~\ref{Problem_presentation}) in cardiac CT and MRI images.  
Our multi-class WHS framework is based on the proposed non-rigid registration pipeline for atlas construction (see in subsection~\ref{Atlas_Construction}), utilizing a multi-resolution strategy to obtain the highest possible mutual information between moving and fixed CT or MRI images. Different parameters of the registration algorithm in our pipeline, such as optimizer, interpolator, metric, resampling technique, and transformation, are tuned for achieving better spatial alignment between moving and fixed CT or MRI images.
We also develop various fusion strategies of multiple atlases to delineate the different anatomical structures for ablation studies.  We further incorporate non-rigid registration into expectation-maximization for the WHS to compare with the proposed statistical segmentation algorithm. Besides, we have implemented other CNN-based supervised methods, where we employ three variants of encoder-decoder networks. The best performing CNN network is compared with the proposed statistical pipeline.  We validate all the extensive experiments utilizing the publicly available dataset named MM-WHS-2017 (see details in subsection~\ref{dataset}). The submitted pipeline exceeds state-of-the-art results for the WHS on the used dataset to our most trustworthy knowledge.

The rest of the sections are manifested as follows: section \ref{MaterialsandMethods} illustrates the used datasets to bear extensive experiments and the different methodologies.
Section \ref{ResultsandDiscussion} describes the obtained results accompanying with a precise analysis and state-of-the-art connections.
Finally, section \ref{Conclusions} terminates the paper with prospective future acts.

\section{Materials and Methods}
\label{MaterialsandMethods}
This section elaborates on the materials and methodologies in the article. We present the utilized dataset and our proposed WHS schemes in subsections \ref{dataset} and \ref{Methodologies}, respectively. Subsections \ref{CNN_based-Methods} and \ref{Evaluation} respectively explain other implemented CNN-based methods for WHS and the hardware \& metrics used to evaluate the experimentation.

\subsection{Dataset}
\label{dataset}
All the comprehensive experiments were conducted utilizing publicly available MM-WHS-$2017$ dataset \cite{zhuang2019evaluation}, as it is commonly used in recent articles (see in Table~\ref{tab:different_Methods}), which contains $40$ cardiac whole heart volumetric MRI and CT data.
The VOIs include the seven different substructures in the utilized WHS dataset, as described earlier in subsection~\ref{Problem_presentation} (see in Table~\ref{tab:whs_sub}). 
The different substructures, such as LV, RV, LA, RA, Myo, AO, and PA (see in Fig.~\ref{fig:dataset}) of both the CT and MRI are labeled as $500$, $600$, $420$, $550$, $205$, $820$, and $850$, respectively. 
We aim to segment those seven organs from both the cardiac CT and MRI scans. Hence, we have termed it a multi-class (8-classes) segmentation task, including the background and seven different heart substructures. All the experimental WHS approaches are assessed following a leave-one-out evaluation strategy \cite{gubern2011multi}.

The $20$ CT and $20$ MRI sequences were collected from a $64$-slice CT scanner (Philips Medical Systems, The Netherlands) and a $1.5$T clinical scanner (Philips Healthcare, The Netherlands), respectively. The volumes from different scanners were stored as NIfTI file format in differing image properties and resolutions to provide imperfect training data to promote more robust algorithms' construction \cite{liao2020mmtlnet}. The former volumetric CT images were obtained in axial view, incorporating the entire heart from the topmost abdominal to the aortic arch with an in-plane resolution of $0.44 \times 0.44$ mm, and the standard slice thickness of $0.60$ mm. The latter volumetric MRI images were accumulated with a resolution of $2 \times 2 \times 2$ mm and reconstructed to nearby $1 \times 1 \times 1$ mm.

\subsection{Proposed Method}
\label{Methodologies}
The WHS's proposed method essentially consists of two integral parts, such as \textit{Atlas construction} and \textit{segmentation strategy}, where we integrate different strategies to select the best performer for WHS. The elaborate discussion of these two crucial parts of our proposed framework is manifested in the following two subsections, \ref{Atlas_Construction} and \ref{Segmentation_Strategies}, respectively.

\subsubsection{Registration and Atlas Construction}
\label{Atlas_Construction}
\begin{figure*}[!ht]
  \centering
  \subfloat{\includegraphics[width=16.01cm, height=7.88cm]{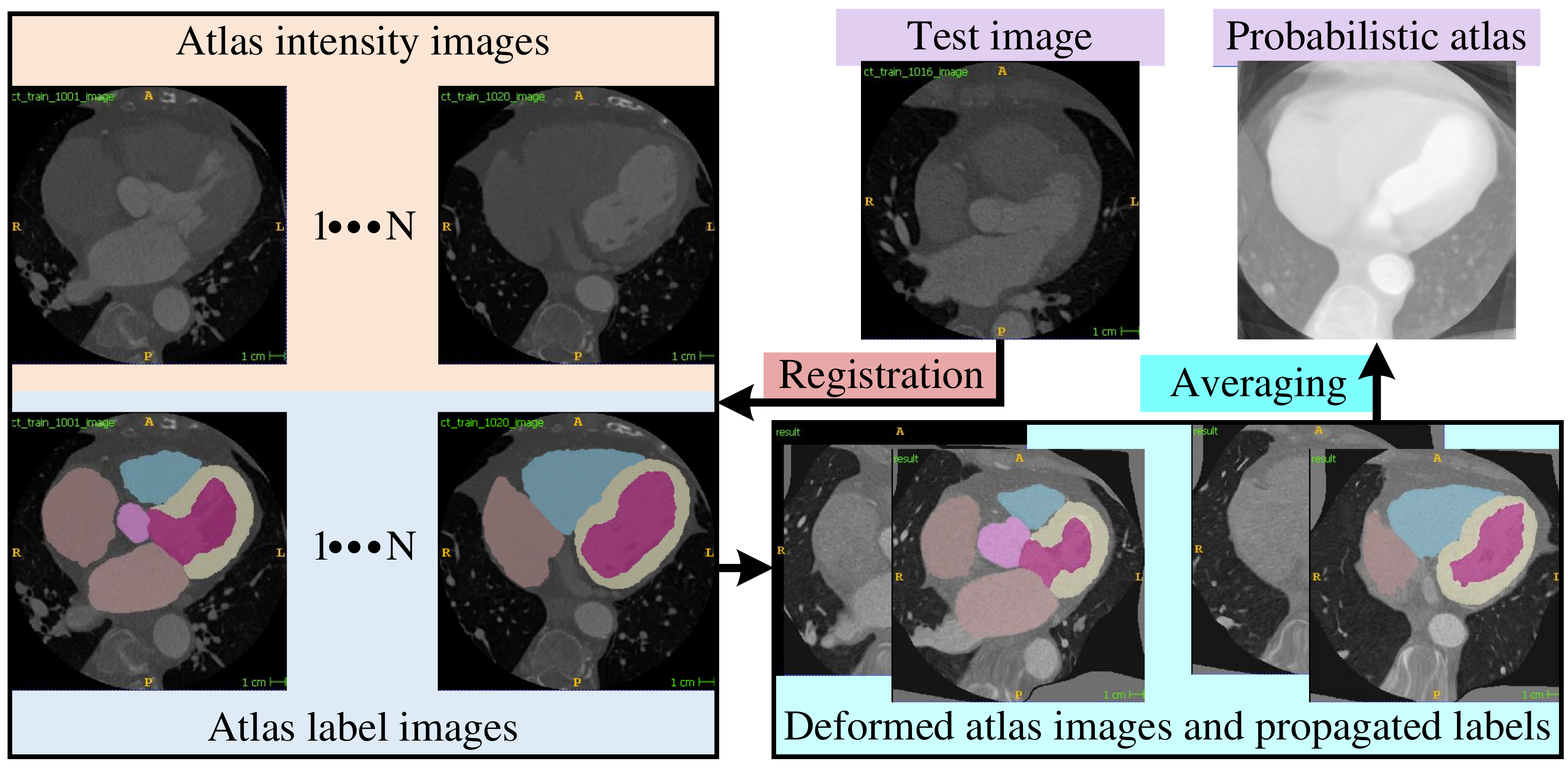}} 
  \caption{Illustration of probabilistic atlas-based segmentation for the whole heart segmentation employing different fusion strategies, where we use non-rigid registration for atlas construction.}
  \label{fig:Probabilistic}
\end{figure*}

Fig.~\ref{fig:Probabilistic} demonstrates the proposed pipeline for atlas-based WHS, where the registration is the crucial integral step to deform a moving CT or MRI image to align with a fixed CT or MRI image spatially.
Let us consider that $I_T$ is an unseen fixed cardiac image (CT or MRI) to be segmented ($L_T$). 
$\{(I_i,L_i)|i=1 ,..., N\}$ is a set of moving images to build atlases, where $I_i$, $L_i$, and $N$ are the $i^{th}$ intensity image, its corresponding $i^{th}$ label image, and the number of member images in the reference volumes, respectively. 
We perform moving-to-fixed registration (non-rigid) to construct a deformed set, $\{(I_i{^D},L_i{^D})|i=1,...,N\}$, where $I_i{^D}$ and $L_i{^D}$ are the resultant deformed intensity image and its corresponding deformed label image.
The deformed intensity image ($I_i{^D}$) and its corresponding deformed label image ($L_i{^D}$) are averaged to construct probabilistic atlas ($I_A$) and probabilistic label ($L_A$), respectively, using \eqref{eq:eq00} and \eqref{eq:eq000}, respectively.  
\begin{equation}
\label{eq:eq00}
\begin{aligned}
    I_A (x,y,z) = \frac{1}{N}\sum_{i=1}^N I_i{^D}(x,y,z) \\ = \frac{1}{N}\sum_{i=1}^N I_i(T_{NR_i}(x,y,z)),
    \end{aligned}
\end{equation} 
\begin{equation}
\label{eq:eq000}
\begin{aligned}
    L_A (x,y,z) = \frac{1}{N}\sum_{i=1}^N L_i{^D}(x,y,z) \\ = \frac{1}{N}\sum_{i=1}^N L_i(T_{NR_i}(x,y,z)),
    \end{aligned}
\end{equation} 
\noindent where $T_{NR_i}$ is a transformation between $i^{th}$ pair of $I_i$ and $I_T$. For $M$ voxels and $k\in\{1, 2, ... ,K=7\}$ target substructure regions, both the $I_A (x,y,z)$ and $L_A (x,y,z)$ probabilistic atlas is a matrix with $N\times K$ elements, and each element ($P_{nk}\in(I_A\, or\, L_A)$) represents the anatomical knowledge about the heart provided by training samples on the prior probability of $n^{th}$ voxel belonging to a particular tissue class $k$.

The intensity image ($I_i$) is appointed as the moving image for deforming to the unseen fixed images ($I_T$) using a non-rigid transformation ($T_{NR}$) \cite{rueckert1999nonrigid, maurer1998investigation, lorenzo2002atlas}. The algorithm of $T_{NR}$ is a combination of a global and local transformations as $T_{NR} (x,y,z) = T_{global} (x,y,z)+T_{local} (x,y,z)$. The global ($T_{global}$) is an affine transformation allowing scaling, translation, rotation, and shearing of $I_i$, whereas the local ($T_{local}$) is a free-form deformation model based on B-splines \cite{eilers1996flexible}. However, in our pipeline, we formulate the registration as an optimization problem in~\eqref{Transformation} for maximizing the similarity between $I_i$ and $I_T$ employing the $T_{NR}$. 
\begin{equation}
\overline{T_{NR}} = \argmin_{T_{NR}} \Upsilon [T_{NR};I_T(x,y,z),I_i(x,y,z)],
\label{Transformation}
\end{equation}
\noindent where $\overline{T_{NR}}$ is an optimal transformation to spatially align $I_i(T_{NR}(x,y,z))$ to $I_T$. $\Upsilon$ is a cost function, which we minimize by designing a pipeline combining different algorithms. The cost function ($\Upsilon$) is optimized using adaptive stochastic gradient descent optimizer \cite{klein2009adaptive}. Other crucial algorithms in our proposed pipeline, such as interpolator, metric, and resample-interpolator, are the B-Spline algorithm with an order of $1$, mutual information \cite{mattes2003pet} with histogram bins of $32$, and B-Spline algorithm with an order of $3$, respectively. A multi-resolution strategy, with resolutions of $4$, is used to bypass local minima \cite{isgum2009multi}. 
We use \textit{Elastix-$5.0$}\footnote{\url{https://elastix.lumc.nl/doxygen/index.html}} \cite{klein2009elastix} to implement our proposed registration pipeline.  
 In this work, a full probabilistic atlas is built and evaluated following a leave-one-out evaluation strategy \cite{gubern2011multi}.

\subsubsection{Segmentation Strategies}
\label{Segmentation_Strategies}
This subsection presents and designs several integral strategies in our proposed pipeline (see in Fig.~\ref{fig:Probabilistic}) to delineate different anatomical substructures (see in Table~\ref{tab:whs_sub}); once the moving images are deformed to a fixed image employing a non-rigid registration. 

\paragraph{Multi-atlas Label Propagation}
The easiest and quickest technique to assign a label to each voxel of an input test image is the label propagation of deformed labels ($L_i{^D}$) to the unseen test image space \cite{klein2005mindboggle, ciofolo2009atlas, wu2006automated}. Multi-atlas Label Propagation (MALP) trades better with the registration errors comparing a single atlas and better accounts for anatomical variability \cite{cabezas2011review}. The MALP also strengthens over single label propagation, as it can reject outliers (minority) of the deformed labels ($L_i{^D}$). The Majority Voting Fusion (MVF) \cite{zhuang2016multi} is a MALP approach, which counts the number of atlases provided the same label for a test voxel, $n$, in the same spatial location. The resultant segmented VOI ($L_T$) can be estimated from $L_i{^D}$, as in \eqref{eq:eq1}: 
\begin{equation}
\label{eq:eq1}
    L_T(n) = \argmax_{k\in\{1, ... ,K\}}{\sum_{i=1}^{N} \Gamma (L_i{^D}(n),k)},
\end{equation}
\noindent where $\{1, 2, ..., K\}$ is a set of K (=$7$) labels of the heart anatomy (see in subsection~\ref{dataset}), $\Gamma (L_i{^D}(n),k)$ is a counting function, as defined in~\eqref{cf}.
\begin{equation}
\Gamma (L_i{^D}(n),k) = 
    \begin{cases}
  1 & \text{if $L_i{^D}(n)=k$},\\
  0 & \text{if $L_i{^D}(n)\neq k$}.
\end{cases}
\label{cf}
\end{equation}

\noindent The other method of MALP is the estimation median value of all the candidate labels in a particular voxel location. The median value of each test voxel ($n$) from the deformed labels ($L_i{^D}$) also provides a robust estimation of resultant label ($L_T$) \cite{amiri2014machine}, which can be formulated as 
$L_T(n) = Median (L_k{^S}(n)), \forall k\in K$, where $L_k{^S}(n),\forall k\in K$, is a stacked (in $4^{th}$ dimension) deformed label of $L_i{^D},\forall i\in N$, which has $N$-values for all voxels, $n$. 

\paragraph{Probabilistic Atlas-based Segmentation (PAS)}
In our propsoed PAS framework, we define the actual moving labels as $X$ and the target fixed image as $Y$, where the components of $X$ and $Y$ are prepared by a spatial location expressed by $n\in J$, wherever $J$ is the simplistic $3D$ rectangular grid index $(x, y, z)$.  Let us considering that $X=(x_1,x_2,...,x_M)$, $Y=(y_1,y_2,...,y_M)$, and $A=(a_{1},a_{2},...,a_{M})$ are the sample realizations of label image, intensity image, and probabilistic  label, respectively, where $M$ is the total voxel number. Example space of $X$ is indicated by $\Omega_x$, where $\Omega_x = \{x : x_n \in\{k=1,2,...,K\}, \forall n \in J\}$. The probability atlas is $K$-vector $a_{j} = (a_{j1}, a_{j2},...,a_{jK}),\forall j\in M$, where all the ingredient corresponds to expectation of $K$-different heart substructures. The prior probability for all voxels can be expressed as $P(x_n=k)=a_{nk},\forall n \in J; k\in K$.

\begin{figure*}[!ht]
  \centering
  \subfloat{\includegraphics[width=16.1cm, height=8.62cm]{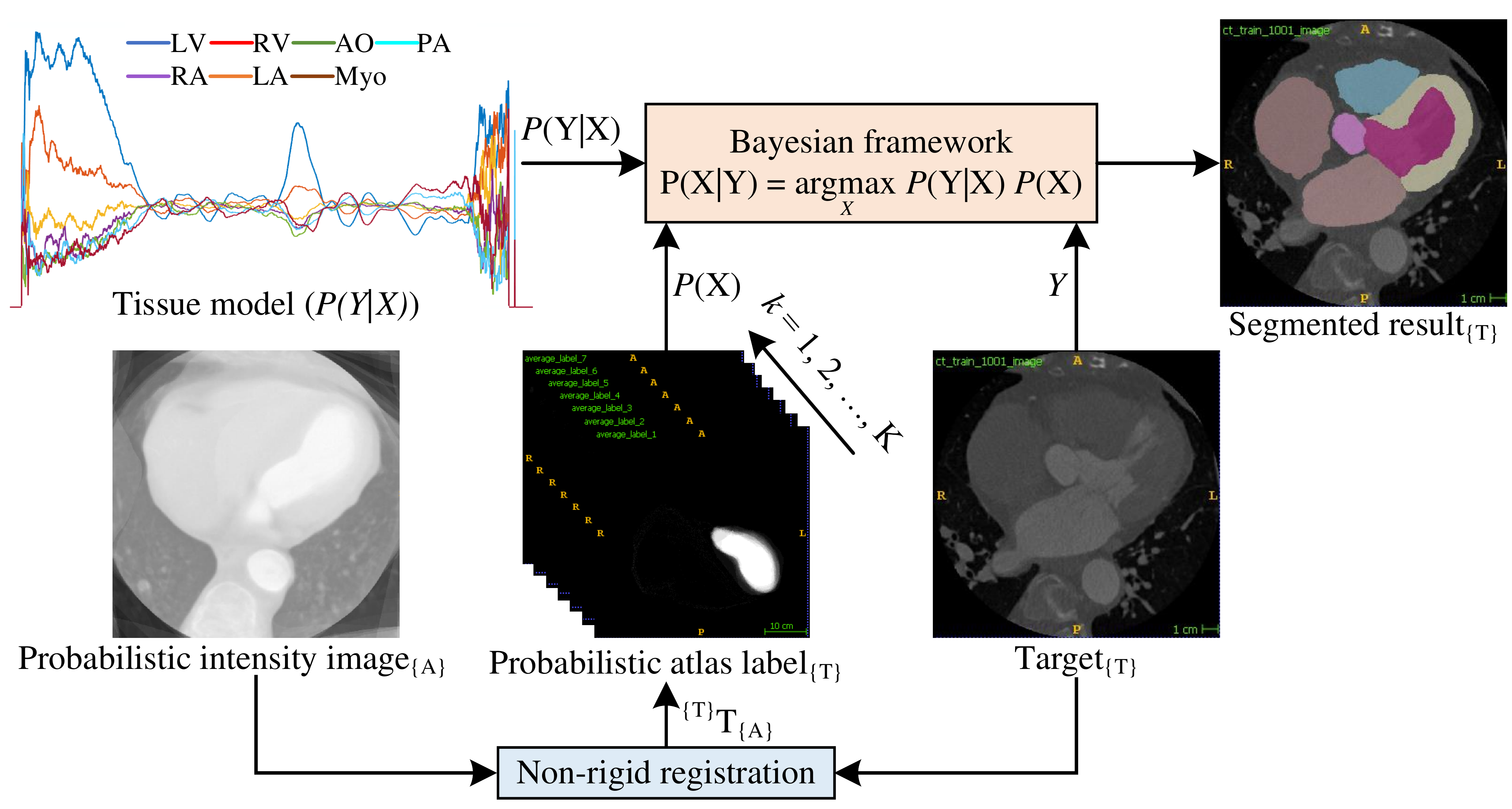}} 
  \caption{Voxel classification algorithm overview for the WHS, where the probabilistic atlas labels are transformed to the unseen test image space $\{T\}$ utilizing the atlas' anatomical image. The probabilistic atlas, also known as the tissue models, and the unseen objective image are given to the Bayesian interface as a prior probability $P(X)$, conditional probability $P(Y|X)$ and set of intensity values $Y$, respectively. The Bayesian interface estimates the posterior probability for the segmentation of given $X$ maximizing the $P(X)\cdot P(Y|X)$.}
  \label{fig:PAS}
\end{figure*}

The hindrance comprises determining the label $X$ that adequately illustrates the provided observation $Y$ according to any loss function. As a decision rule, we chose MAP (maximum a posteriori) and the formula for the realization of estimating of $X$ as $\hat{x} = \argmax_x{P(X=x|Y=y)}$. 
The posterior probability ($P(X|Y)$) can be written as the multiplication of probability distribution ($P(Y|X)$), also named tissue model \cite{park2003construction,gubern2011multi}, and prior probability ($P(X)$), according to the Bayes theorem. 
$P(Y|X)$ is defined by signal intensity tissue models immediately formed from the scans and hand-operated segmentation of the data set. An intensity value's histogram is constructed for each heart substructure considering the given volumes' voxels, which belong to it, using hand-operated segmentation. 
In this work, we estimate the probability distribution $P(Y|X)$ of the given image $Y$ for the provided appropriate segmentation $X$ from training data. Algorithm~\ref{alg:tissue_model} shows detail process of estimating $P(Y|X)$ using the number of bins ($N_b$) for the histograms as $4096$.
\begin{algorithm}[!ht]
\begin{algorithmic}[1]
\STATE Initialize the counters as $C_b^k=0,\forall b \in N_b; k\in K$
\FOR{all training images}
\STATE Rescale the image ([$0\sim N_b-1$]) and extract VOI
\STATE Accumulate $N^k,\forall k \in K$ with corresponding pixel numbers.
\STATE Compute histograms $H^k,\forall k \in K$, and accumulate in~counters $C^k += H^k, \forall k \in K$
\ENDFOR
\STATE Normalize histograms $C^k/N^k, \forall k \in K$
\STATE Scale the histograms as $C^k_b = \frac{C^k_b}{\sum_{k=1}^K C^k_b},\forall b \in N_b; k\in K$, so that $\sum_{k=1}^K C^k_b=1,\forall b \in N_b$
\end{algorithmic}
\caption{Estimation of $P(Y|X)$ of the image $Y$ for given segmentation $X$.}
\label{alg:tissue_model}
\end{algorithm}
Fig.~\ref{fig:PAS} exhibits an illustration of the Bayesian voxel classification algorithm consolidating the application of the probabilistic atlas of the WHS.
Besides, the probabilistic atlas gives the probability distribution $P(X)$ once it has been mapped onto the target space employing the corresponding registration method utilized in its creation.

\paragraph{Expectation Maximization (EM)}
The EM algorithm is an iterative approach that classifies data points with the Gaussian Mixture Model (GMM), updates model parameters with the newly classified data, and classifies data points with the new parameters. In our implementation, we use EM to learn the parameters of the GMM in an unsupervised fashion for the segmentation of different heart substructures from the whole cardiac CT or MRI images. 
Let us assume that $X=(x_1,x_2,...,x_M)$ is an unseen target CT or MRI image, where $x_i$, $\forall i \in M$, is $d$-dimensional voxel intensities and $M$ is total voxel numbers. Suppose that $x_i,\forall i \in M$, is formed in an IID order from an underlying density of GMM model ($P(X)$) with $K$ ingredients, as in~\eqref{gmm}. 
\begin{equation}
    P(X|\Theta) = \sum_{k=1}^{K} \alpha_k P_k(X|Z_k,\theta_k), 
    \label{gmm}
\end{equation}
\noindent where $P_k(X|Z_k,\theta_k)$, $\alpha_k$, and $Z_k$ are the $k^{th}$ ingredient of GMM with parameters $\theta_k$, mix-up weights describing the probability that a randomly chosen $X$ is 
formed by $k^{th}$ element, and a vector of $K$ pointer variables that are jointly independent, respectively, where $\sum_{k=1}^K \alpha_k =1$. 
Therefore, the entire set of parameters for a GMM with $K$ elements is  $\Theta =\{\alpha_1,...,\alpha_K, \theta_1,...,\theta_K\}$. We can model $P_k(X|Z_k,\theta_k)$ as Gaussian density function as in~\eqref{gdf} with the parameters of $\theta_k = \{\mu_k,\Sigma_k\}$.

\begin{equation}
 P_k(X|\theta_k) = \frac{1}{(2\pi)^{d/2}|\Sigma_k|^{1/2}} \cdot e^{(-\frac{1}{2}(X-\mu_k)^T\Sigma_k^{-1}(X-\mu_k))}, 
 \label{gdf}
\end{equation}
\noindent where $\forall k\in K$. The parameters in the above equation, such as $\mu_k$ and $\Sigma_k$, respectively denote the mean and co-variance or standard deviation of $k^{th}$ Gaussian density function, $P_k(X|\theta_k)$. From the known GMM parameters and using the Bayes rule, the membership probabilities of a given observed vector ($x_i \in X$) can be written, as in~\eqref{membership}:
\begin{equation}
    W_{ik}= \frac{\alpha_k \cdot P_k(x_i|\theta_k) }{\sum_{j=1}^{K} \alpha_j \cdot P_k(x_i|\theta_j)}, \forall k \in K, 
    \label{membership}
\end{equation} 
\nonumber where $\sum_{k=1}^{K} W_{ik}=1$. 
In our pipeline, we propose to initial parameters of the EM algorithm using the probabilistic atlas. The convergence of the EM algorithm is recognized by estimating the log-likelihood value, as in~\eqref{likelihood}, after each iteration and stopping when it seems not to increase significantly from one iteration to the succeeding. 
\begin{equation}
\label{likelihood}
\begin{aligned}
    log(l(\Theta)) = \sum_{i=1}^M log(P(X|\Theta)) \\ = \sum_{i=1}^M \bigg( log \Big(\sum_{k=1}^K \alpha_k \cdot P_k(X|Z_k,\theta_k) \Big) \bigg) .
    \end{aligned} \tag{9}
\end{equation} 
The optimization of the EM algorithm in our pipeline essentially includes the following two steps to obtain optimal parameters of the GMM. 

\textbf{Expectation-step (E-step):} It includes the following three steps. 
\begin{enumerate}[Step 1:]
\item Denote the current parameter values (such as $\alpha_k$, $\mu_k$, and $\Sigma_k$, $\forall k \in K$) as $\Theta$, where $\sum_{k=1}^K \alpha_k =1$.  

\item Compute membership probabilities ($W_{ik}$) having the size of $M\times K$ using the equation mentioned earlier in~\eqref{membership}, $\forall x_i \in X$, $1\leq i \leq M$, $\forall k \in K$, where $\sum_{k=1}^K W_{ik} =1$.   

\item Compute the log-likelihood ($L$) using \eqref{likelihood} for the current parameters ($\Theta$), which is termed as $L_{old}$.
\end{enumerate}

\textbf{Maximization-step (M-step):} The matrix of membership weights ($W_{ik}$) is used to update the parameters $\Theta$ using the following three steps. 
\begin{enumerate}[Step 1:]
\item Update the mixture weights ($\alpha_k$, $\forall k \in K$) by using $\alpha_k^{new}=\frac{1}{M}\sum_{i=1}^M W_{ik}$, where $\sum_{k=1}^K \alpha_k =1$.    

\item Update parameters, $\theta_k = \mu_k, \Sigma_k$, $\forall k \in K$, by using the following equations, 
\begin{equation}
\begin{aligned}
\mu_k^{new} = \frac{1}{\Acute{M}_k} \sum_{i=1}^M W_{ik}\cdot x_i, \forall k \in K,
  \nonumber
\end{aligned}
\end{equation}

\begin{equation}
\begin{aligned}
\Sigma_k^{new} = \frac{1}{\Acute{M}_k}\sum_{i=1}^M W_{ik} \cdot \big(x_i-\mu_k^{new}\big)\big(x_i-\mu_k^{new}\big)^T,
  \nonumber
\end{aligned}
\end{equation}

where
\begin{equation}
\begin{aligned}
\Acute{M}_k = \sum_{i=1}^M W_{ik}, \forall k \in K.
  \nonumber
\end{aligned}
\end{equation}

\item Compute the log-likelihood ($L$) using \eqref{likelihood} for the updated parameters ($\Theta_{new}$), which is termed as $L_{new}$.
\end{enumerate}
We run the EM algorithm until the convergence, which means we repeat until there is no significant change between $L_{old}$ and $ L_{new}$. Finally, the membership probabilities ($W_{ik}$) are used to delineate WHS's different substructures.  

\paragraph{PAS+EM}
In our proposed PAS+EM algorithm, we use posterior probabilities ($P(X|Y)$) and membership probabilities ($W_{ik}$) from PAS and EM algorithms, respectively, to delineate the different structures for WHS. For doing so, we use the following equation, as in~\eqref{pasem}, to obtain the probabilities for our PAS+EM algorithm. 
\begin{equation}
P_{PASEM} = W_{ik}\cdot P(X|Y), \forall k \in K, \tag{10}
\label{pasem}
\end{equation}
\nonumber where all $P_{PASEM}$, $W_{ik}$, and $P(X|Y)$ are $M\times K$ matrix. Finaly, in the end, we use $P_{PASEM}$ to delineate the different structures for WHS.

\subsection{CNN-based Methods}
\label{CNN_based-Methods}
We have implemented a CNN-based semantic segmentation method besides our proposed atlas-based segmentation method to perform comprehensive ablation studies.  

\begin{figure}[!ht]
  \centering
  \subfloat{\includegraphics[width=8cm, height=5cm]{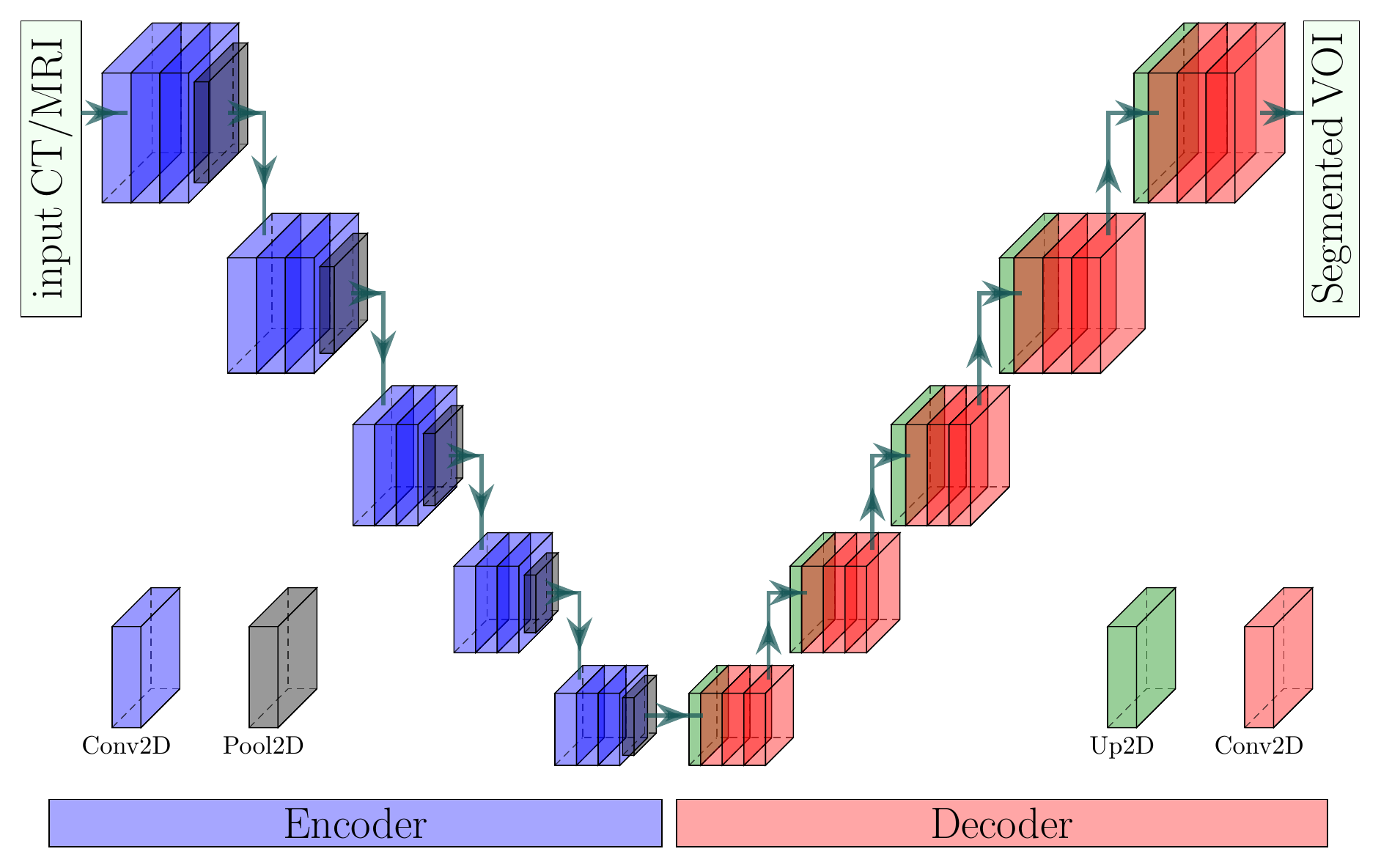}}
  \caption{A pictorial presentation of EDNet applying a pre-trained VGG-16 network as an encoder for transferring the previous ImageNet knowledge to our WHS task. The same number of pooling and upsampling layers are used in the encoder and decoder to regain the output's input resolutions.}
  \label{fig:EDNet}
\end{figure}

Segmentation of medical image has attained enormous progress notably since $2015$ after the proposing of UNet \cite{ronneberger2015u}. Currently, CNN-based networks have been extensively practiced for the medical imaging domain, exceeding conventional image analysis techniques relying on hand-crafted features \cite{tajbakhsh2020embracing}. However, the CNN-based network for segmentation incorporates two fundamental elements: the encoder and the decoder \cite{ronneberger2015u}. An encoder consists of convolutional and pooling layers. The convolutional layers produce feature maps, whereas the pooling layers continuously decrease these feature maps' dimension to gain more critical features with more eminent spatial invariance \cite{hasan2020dsnet}. 
The decreased resolution feature maps also enlarge the maps' field-of-view and diminish the computational expense \cite{long2015fully}. 
The decoder projects the distinctive lower resolution features discovered by the encoder onto the higher resolution pixel space to achieve a compact pixel-wise labeling \cite{garcia2018survey}.
However, the simple encoder-decoder network, named EDNet, in our implementation is depicted in Fig~\ref{fig:EDNet}.
The encoder in EDNet is a VGG-16 \cite{simonyan2014very} with the pre-trained weight on ImageNet \cite{deng2009imagenet},
which has five Convolutional Blocks (CB) and thirteen convolutional layers. Each CB's output is an input to the next CB through a pooling layer with a stride of $2\times 2$. Hence, the encoder's output feature map has a resolution of $m/2^5\times n/2^5$ for an input resolution of $m\times n$.
However, the decoder has five blocks to obtain the input resolutions of the output WHS masks ($m\times n$), where we apply 2D upsampling, with a stride of $2\times 2$, convolution with a kernel of $3\times 3$, and a batch normalization \cite{ioffe2015batch} in each decoder block.

\begin{figure}[!ht]
  \centering
  \subfloat{\includegraphics[width=8cm, height=5cm]{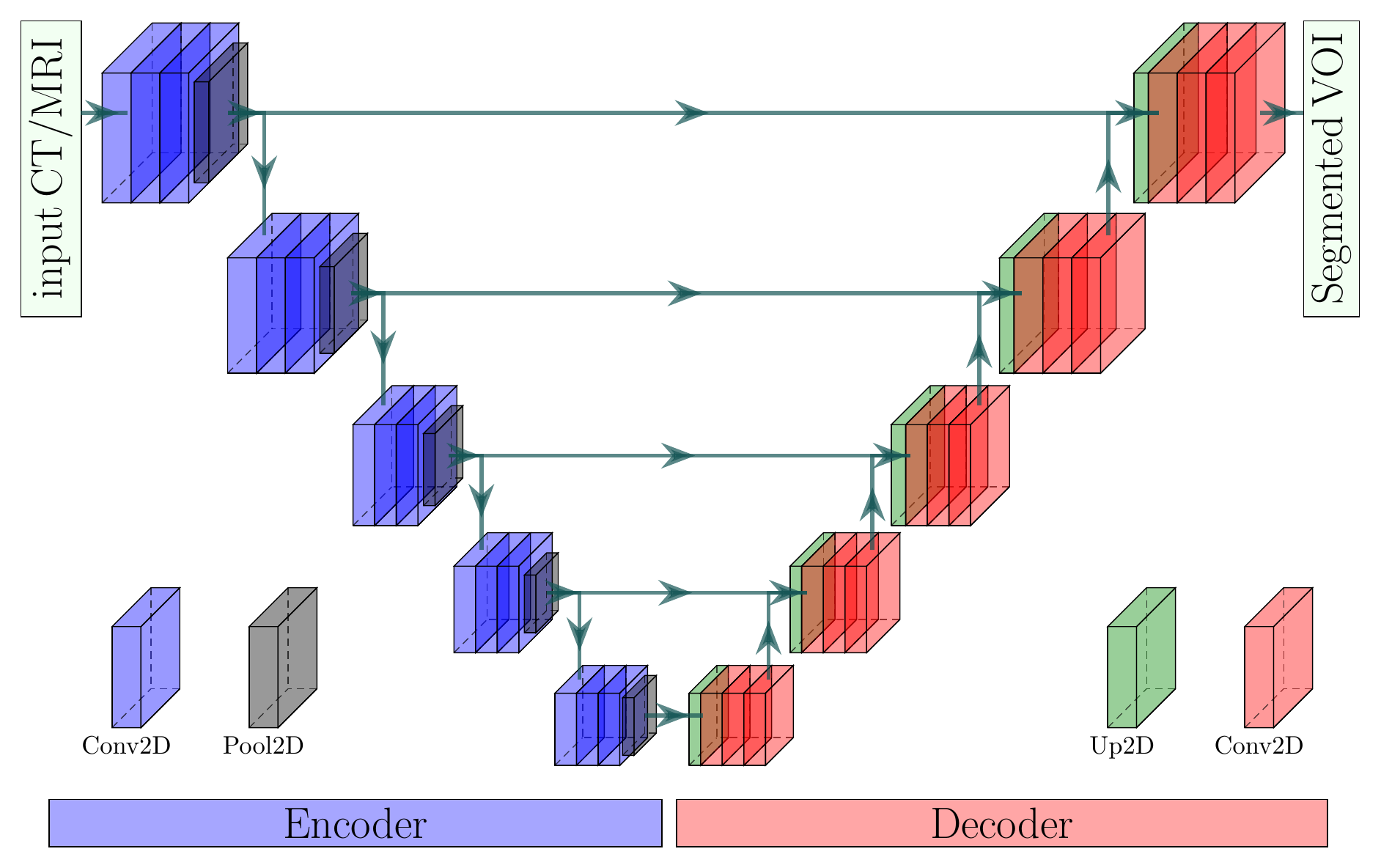}}
  \caption{The architecture of our VGG-UNet, where we concatenate the encoder's features with the same scale features of the decoder to compensate for the lost spatial information in the encoder. The same number of pooling and upsampling layers are used in the encoder and decoder to regain the output's input resolutions.}
  \label{fig:UNet}
\end{figure}

However, the decreased feature maps due to pooling undergo spatial knowledge elimination injecting roughness, poor border knowledge, checkerboard artifacts, over-, and under-segmentation in the segmented substructures \cite{hasan2020dsnet, long2015fully, ronneberger2015u, odena2016deconvolution}. 
To overcome these problems, the authors in \cite{ronneberger2015u} introduced skip connections in a UNet, permitting the decoder to retrieve the associated features discovered at all encoder steps that were missed due to subsampling in the encoder. The feature maps from the encoder's antecedent layers are concatenated with the decoder's identical scale through the appliance of skip connections. Applying the skip connection of the popular UNet, we propose a VGG-UNet, where we have also employed the VGG-16 network as an encoder, as shown in Fig.~\ref{fig:UNet}. 
In our VGG-UNet, we apply the skip connections holding ladder-like compositions \cite{rasmus2015semi} motivated by UNet to succeed in the pooling weaknesses. All pooled layers of our network are concatenated channel-wise to a deconvoluted feature map with identical dimensions, where it acts as an offsetting link for the spatial knowledge dropped due to subsampling in the encoder. 

\begin{figure}[!ht]
  \centering
  \subfloat{\includegraphics[width=8cm, height=5cm]{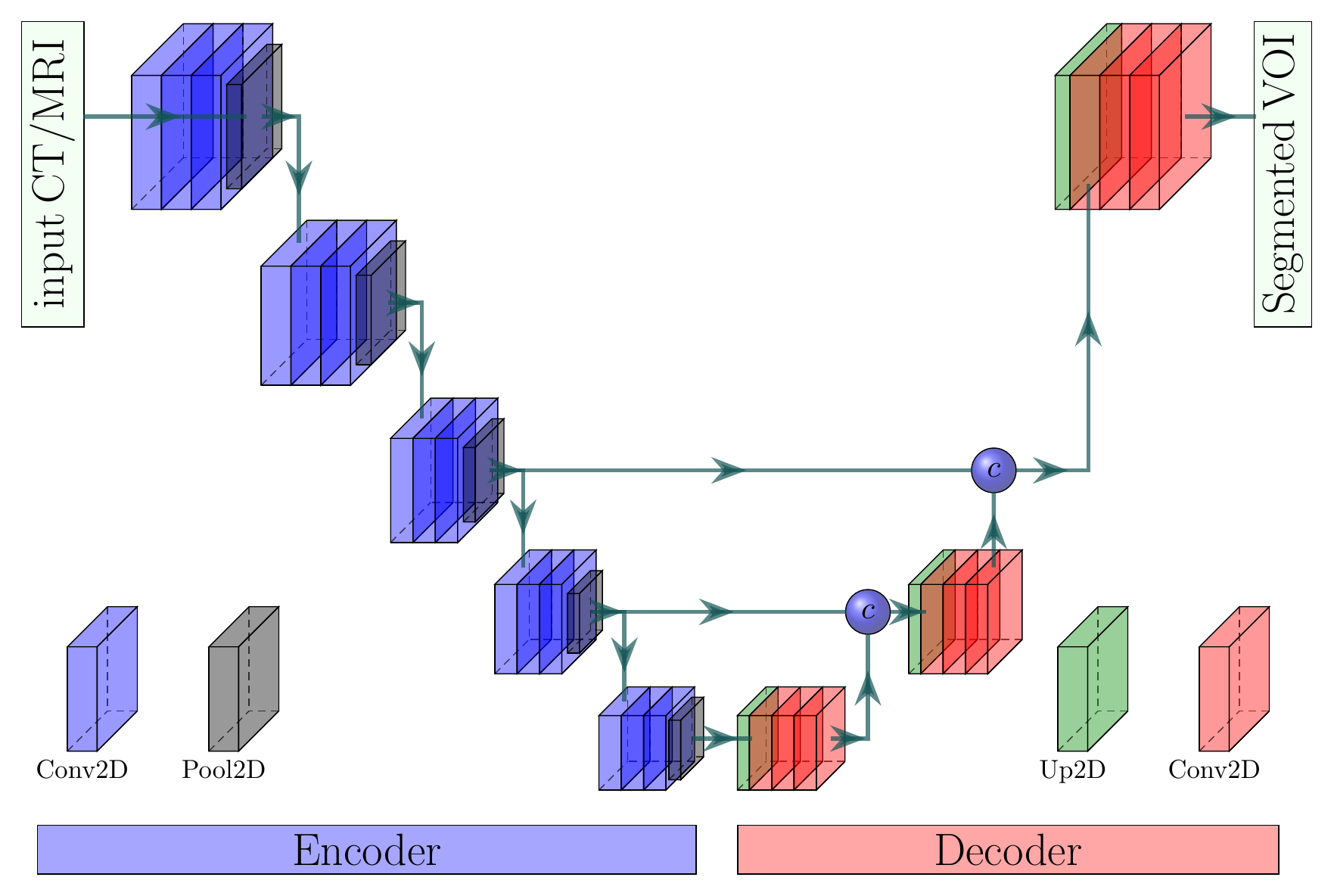}}
  \caption{The structure of our VGG-FCN8s, where we apply a pre-trained VGG-16 as a feature learner. The output WHS masks are obtained from the fused feature maps from the fifth, fourth, and third pooling layers of VGG-16. The same number of pooling and upsampling layers are used in the encoder and decoder to regain the output's input resolutions.}
  \label{fig:FCN8s}
\end{figure}

Again, the authors in \cite{long2015fully} combined features at varied coarseness levels of the encoder in their popular FCN to polish the output segmented VOIs. In this article, we propose a VGG-FCN, as shown in Fig.~\ref{fig:FCN8s}, with the pre-trained VGG-16 network in the encoder. 
The output feature map of such an encoder has $32$-times fewer resolutions as VGG-16 has five pooling layers.
The employment of $32\times$ upsampling can output the segmented mask with the same size as the input image, called FCN32s. Such an upsampling produces a coarse or rough mask as it utilizes only global information from the more in-depth high-level features. 
The in-depth features are achieved while operating deeper, which causes the spatial location information lost. That indicates that output from shallower layers has more location knowledge. If we combine both local and global information, it can enhance the segmentation result. The output from the fifth pooling of VGG-16 is $2\times$ upsampled and fused with the fourth pooling in our network. Then, the combined map is $2\times$ upsampled and again linked with third pooling. Finally, we perform $8\times$ upsampling, which provides a segmented WHS mask. Hence, the proposed VGG-FCN is also named VGG-FCN8s, where we use both the local and global information to obtain the final mask. 

We implemented three semantic segmentation network variants to compare them with the proposed Atlas-based segmentation method of WHS. We apply IoU as an objective metric ($M_{seg}$) in~\eqref{loss} throughout the WHS training for all networks, as mentioned above.

\begin{equation} \label{loss}
M_{seg}(y,\hat{y})= \frac{\displaystyle\sum_{i=1}^{N}y_i\times \hat{y_i}}{\displaystyle\sum_{i=1}^{N}y_i+\displaystyle\sum_{i=1}^{N}\hat{y_i}-\sum_{i=1}^{N}y_i\times \hat{y_i}}, \tag{11}
\end{equation}

\noindent where the parameters, such as $y$, $\hat{y}$, and $N$, respectively indicate the actual label, prognosticated label, and the total voxel numbers. The multiplicative term of $y$ and $\hat{y}$ in the above equation is the estimation of identity (intersection) connecting the actual and predicted VOIs. The binary cross-entropy is utilized as the WHS's cost functions by the networks, such as EDNet, VGG-UNet, and VGG-FCN8s, optimized employing the Adam optimizer \cite{kingma2014adam} with initial Learning Rate ($LR$) and exponential decay rates ($\beta{_1},\,\beta{_2}$) as $LR=0.001$, $\beta{_1}=0.9$, and $\beta{_2}=0.999$, respectively, without AMSGrad variant. The LR is decreased subsequent $5$ epochs by $10.0\,\%$ if validation loss obstructs progressing. The initial epochs are set as $200$, which is suspend the network training utilizing a callback function after the validation loss becomes stagnated.

\subsection{Evaluation}
\label{Evaluation}
The extensive experimentations are conducted on a computer operating on \textit{Windows-10} system with the following hardware configuration:
Intel\textsuperscript{\tiny\textregistered} Core\textsuperscript{\tiny{TM}} i$7$-$7700$ HQ CPU @ $2.80\,GHz$ processor with Install memory (RAM): $32.0\,GB$ and GeForce GTX $1080$ GPU with $8\,GB$ GDDR$5$ memory. The proposed pipeline and CNN-based models are designed utilizing the MATLAB R2020a and Python programming language with various Python and Keras APIs. 

We use mean Volume Overlapping Error (mVOE), mean Sensitivity (mSn), and mean Dice Similarity Coefficient (mDSC) to quantify the segmentation accuracy, which is defined as follows:

\begin{equation}
\begin{aligned}
mVOE=  \frac{1}{S}
\displaystyle\sum_{j=1}^{S} \bigg(1-\frac{TP_{j}}{ TP_{j}+FN_{j}+ FP_{j}}\bigg),
  \nonumber
\end{aligned}
\end{equation}

\begin{equation}
\begin{aligned}
mSn= \frac{1}{S} \displaystyle\sum_{j=1}^{S}\frac{TP_{j}}{TP_{j}+FN_{j}},
  \nonumber
\end{aligned}
\end{equation}

\begin{equation}
\begin{aligned}
mDSC= \frac{1}{S}
\displaystyle\sum_{j=1}^{S}\frac{2\cdot TP_{j}}{2\cdot TP_{j}+FN_{j}+ FP_{j}},
  \nonumber
\end{aligned}
\end{equation}

\noindent where $S$, $TP$, $FN$, and $FP$  respectively denote the number of slices in an unseen test image, true-positive region (VOI as a VOI), false-negative region (VOI as a background), and false-positive region (background as a VOI). mSn determines the percentage of correctly segmented VOI of the true VOI, whereas mVOE and mDSC estimate the difference and similarity between true and segmented VOIs.

\section{Results and Discussion}
\label{ResultsandDiscussion}
This section is dedicated to the presentation of extensive experimental results. In subsection~\ref{Results_on_CT_MRI_scans}, we present and interpret the results of eight different approaches (described in subsections~\ref{Methodologies} \& \ref{CNN_based-Methods}) utilizing the both CT and MRI scans. 
We have grouped those eight distinct methods into three categories: CNN-, MALP-, and probabilistic atlas-base methods, where each group has several methods. Firstly, we perform ablation studies among different methods in each group on the same dataset in subsections~\ref{Results_for_CNN_based_methods}, ~\ref{Results_for_MALP_based_methods}, and ~\ref{Results_for_probabilistic_atlas_based_methods}, respectively.
Finally, in subsection~\ref{Result_comparisons}, we analyze the best methods from all the groups for both the MRI and CT scans, as well as we compare other seventeen different published methods with our best performing method on the same dataset.

\subsection{Experimental Results}
\label{Results_on_CT_MRI_scans}
\begin{figure*}[!ht]
  \centering
  \subfloat[WHS results utilizing CT scans]{\includegraphics[width=8.5cm, height=7cm]{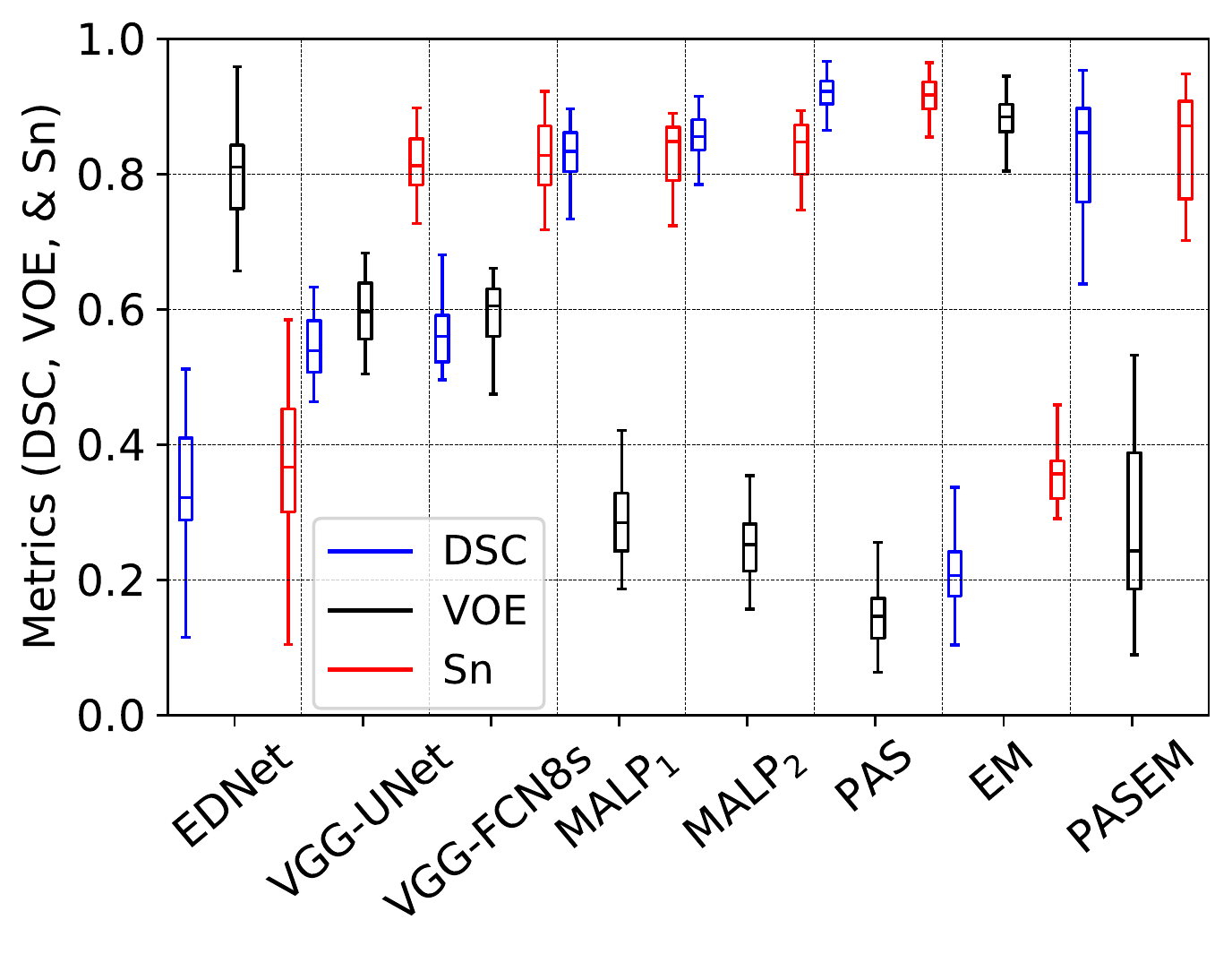}}
  \subfloat[WHS results utilizing MRI scans]{\includegraphics[width=8.5cm, height=7cm]{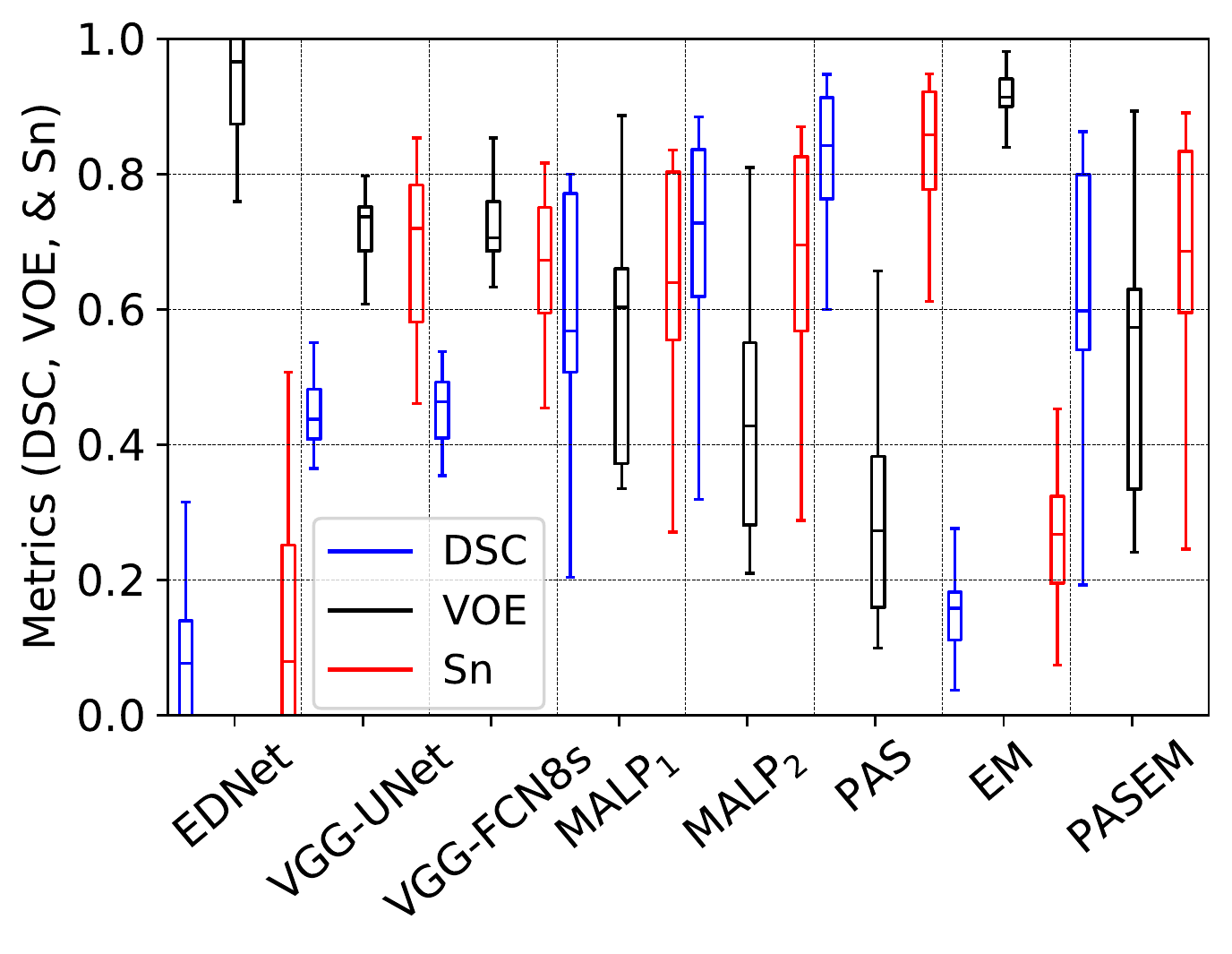}}
  \caption{The Box and Whisker visualization of DSC (blue), VOE (black), and Sn (red) for the WHS utilizing (a) CT and (b) MRI scans employing eight different methods, as described in subsections~\ref{Methodologies} \& \ref{CNN_based-Methods}. The MALP$_1$ and MALP$_2$ indicate a MALP method with the median and MVF schemes, respectively.}
  \label{fig:Box_CT_MRI_qual}
\end{figure*}

\begin{figure*}[!ht]
  \centering
  \subfloat[Qualitative WHS results in CT scans applying different methods]{\includegraphics[width=17.5cm, height=10.3cm]{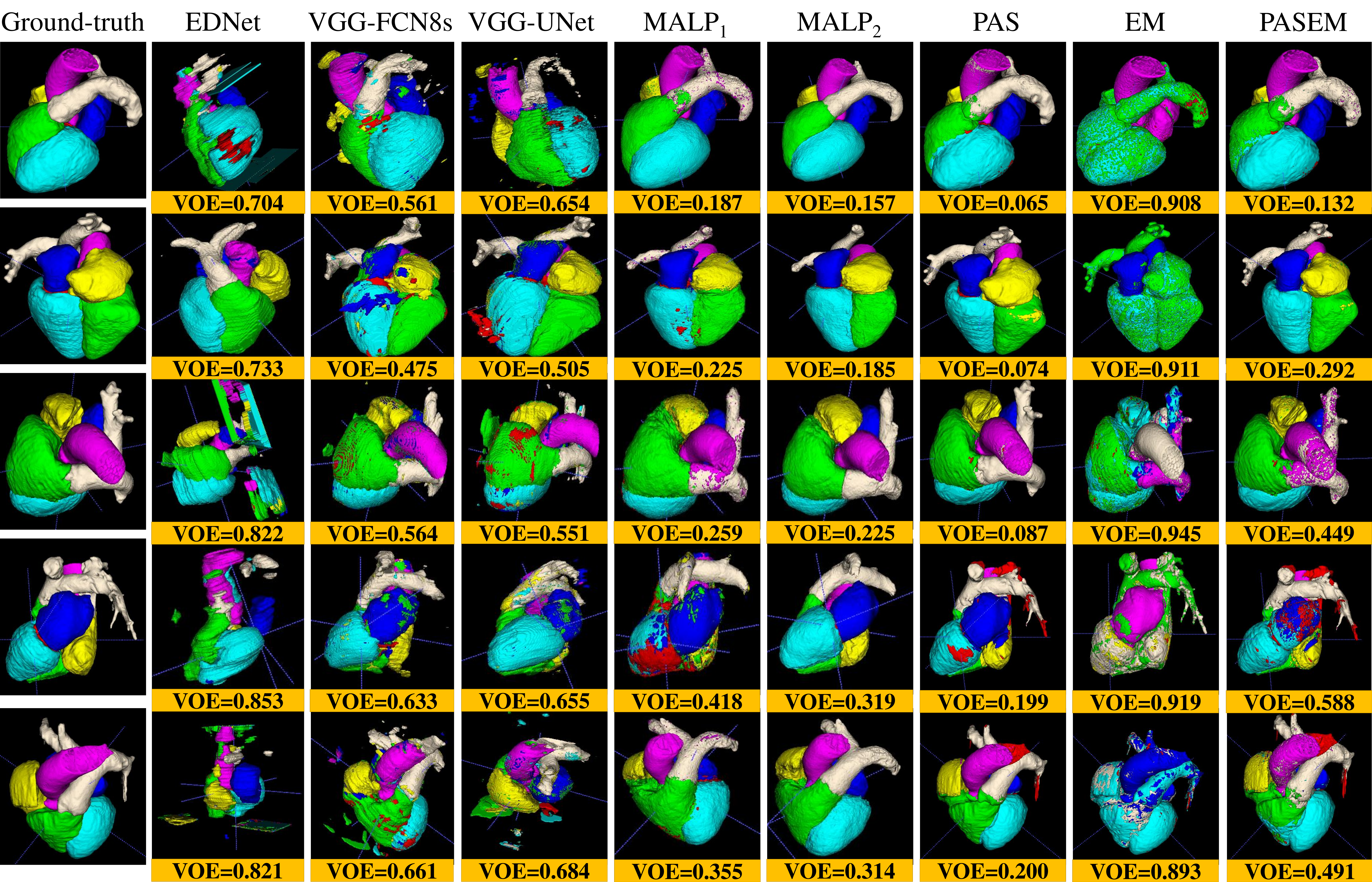}} \\
  \subfloat[Qualitative WHS results in MRI scans applying different methods]{\includegraphics[width=17.5cm, height=10.3cm]{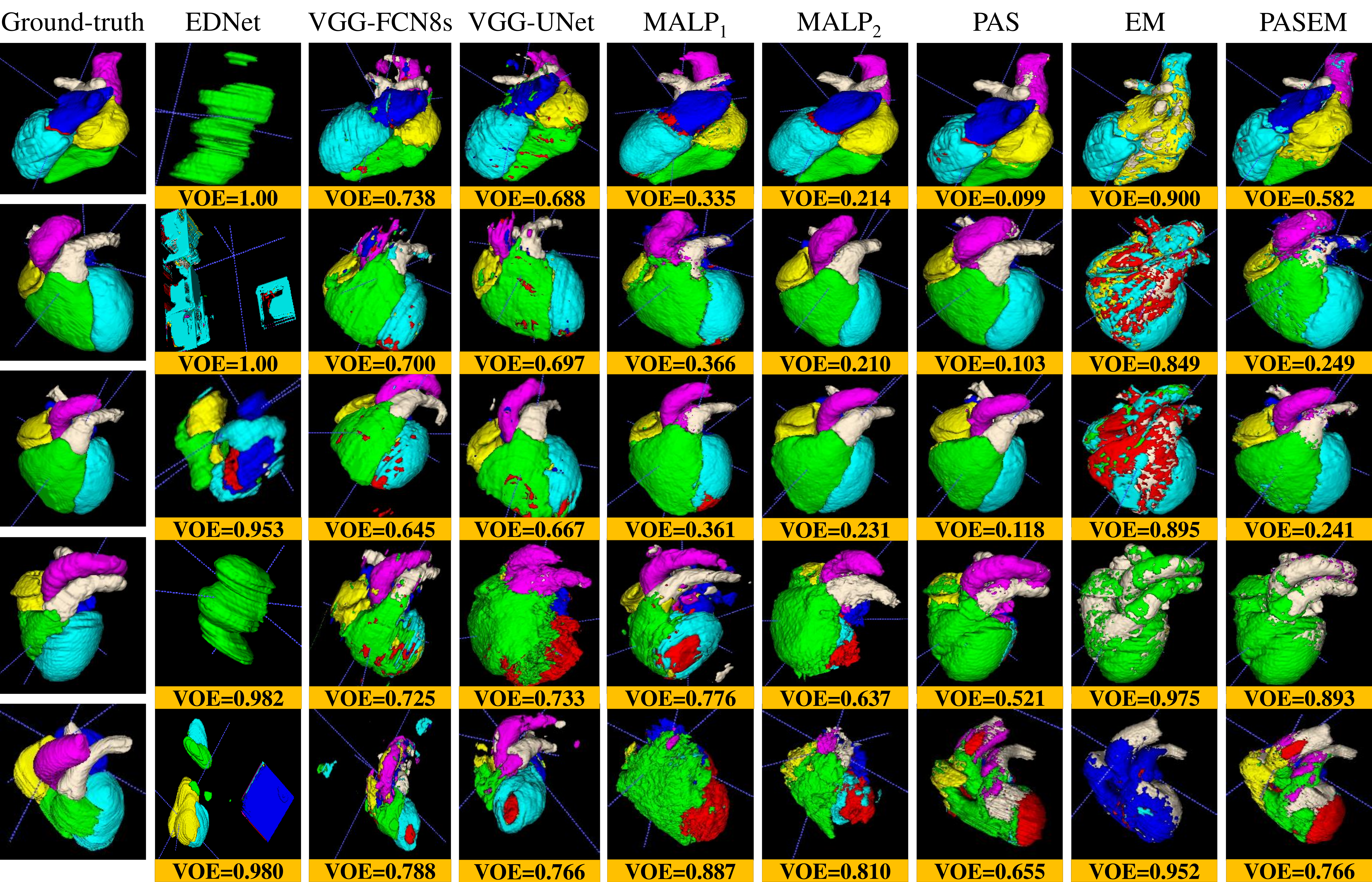}}
  \caption{The qualitative WHS results from our different methods in CT (top) and MRI (bottom) scans, where the first three and last two rows of both the figures are for the best- and worst-performing examples for the best-performing method (PAS), respectively. The MALP$_1$ and MALP$_2$ indicate a MALP method with the median and MVF schemes, respectively.}
  \label{fig:quali_CT_MRI}
\end{figure*}

Table~\ref{tab:CT_MRI_results} demonstrates the obtained results from different methods for the WHS on both the CT and MRI scans explicating different quantitative metrics (see in subsection~\ref{Evaluation}).
\begin{table*}[!ht]
\caption{The quantitative WHS results from our different methods utilizing both the CT and MRI scans. Best metrics are underlined.}
\centering
\begin{tabular}{ll|ccc|ccc}
\hline
\rowcolor[HTML]{C0C0C0} 
\multicolumn{2}{l|}{\cellcolor[HTML]{C0C0C0}}                                    & \multicolumn{3}{c|}{\cellcolor[HTML]{C0C0C0}CT} & \multicolumn{3}{c}{\cellcolor[HTML]{C0C0C0}MRI} \\ \cline{3-8} 
\rowcolor[HTML]{C0C0C0} 
\multicolumn{2}{l|}{\multirow{-2}{*}{\cellcolor[HTML]{C0C0C0}Different methods}} & mVOE           & mSn           & mDSC          & mVOE           & mSn           & mDSC           \\ \hline

\multicolumn{2}{l|}{EDNet}                                                     &   $0.800\pm 0.080$          &   $0.384\pm 0.135$              &  $0.354\pm 0.126$    &   $0.937\pm 0.070$          &   $0.139\pm 0.154$              &  $0.095\pm 0.096$          \\

\multicolumn{2}{l|}{VGG-UNet}                                                       &  $0.596\pm 0.054$           &       $0.796\pm 0.086$       &      $0.543\pm 0.048$    &  $0.720\pm 0.049$           &       $0.693\pm 0.120$       &      $0.444\pm 0.056$          \\ 

\multicolumn{2}{l|}{VGG-FCN8s}                                                       &     $0.590\pm 0.053$        &    $0.828\pm 0.062$        &    $0.564\pm 0.052$       &     $0.719\pm 0.051$        &    $0.654\pm 0.128$        &    $0.450\pm 0.068$                 \\ \hline
                                              & Median                       &  $0.291\pm 0.063$                                                  &       $0.828\pm 0.051$           &          $0.828\pm 0.044$          & $0.556$ $\pm $  $0.169$                                                 &     $0.648$ $\pm $  $0.163$              &             $0.595$  $\pm $ $0.169$   \\

\multirow{-2}{*}{MALP}                           & MVF                          &     $0.251\pm 0.051$                                            &      $0.835\pm 0.046$  &        $0.855\pm 0.033$        &         $0.445\pm 0.180$                                               &          $0.671\pm 0.169$     &            $0.696\pm 0.161$      \\ \hline

\multicolumn{2}{l|}{PAS}                                                         &         $\Cline{0.145\pm0.046}$    &       $\Cline{0.919\pm0.026}$   &  $\Cline{0.921\pm0.027}$                       &         $\Cline{0.291\pm0.162} $    &       $\Cline{0.825\pm 0.126}$   &  $\Cline{0.819\pm0.122} $       \\

\multicolumn{2}{l|}{EM}                                                                     &  $0.878\pm0.040$          &  $0.350\pm0.063$ &               $0.216\pm0.063$       &  $0.917\pm 0.038$          &  $0.260\pm 0.100$ &    $0.152\pm 0.064$          \\ 

\multicolumn{2}{l|}{PAS+EM}                                                      &    $0.331\pm0.144$     &               $0.804\pm0.097$          &    $0.793\pm0.107$                      &    $0.527\pm 0.197$      &              $0.674\pm 0.187$       &   $0.619\pm 0.190$     \\ \hline
\end{tabular}
\label{tab:CT_MRI_results}
\end{table*}
Fig.~\ref{fig:Box_CT_MRI_qual} exhibits the Box and Whisker visualization of all the methods showing the spreads and centers of the DSC, VOE, and Sn metrics for all the 20 test images, either CT or MRI scans.  
Fig.~\ref{fig:quali_CT_MRI} (a) and Fig.~\ref{fig:quali_CT_MRI} (b) display the qualitative segmentation results for both the CT (top) and MRI (bottom), respectively, from all the methods. 

\subsubsection{Results for CNN-based methods}
\label{Results_for_CNN_based_methods}
The experimental results in Table~\ref{tab:CT_MRI_results} quantitatively demonstrate that the EDNet is the worst-performing method than the other two CNN-based methods, such as VGG-UNet and VGG-FCN8s, by the margins of $18.9\,\%$ and $21.0\,\%$ for CT scans and $34.9\,\%$ and $35.5\,\%$ for MRI scans, respectively, concerning the mDSC. Those two networks also defeat the EDNet for the other two metrics (mVOE and mSn) with significant margins (see in Table~\ref{tab:CT_MRI_results}). 
Although those three networks have the same number of convolutional and pooling layers in the encoder and decoder, they constructionally vary in skip connection employment (see details in subsection~\ref{CNN_based-Methods}). 
The results on both the imaging modalities reveal that the appliance of skip connection has outputted better-segmented substructures, as the local information from the shallower layers is utilized to reconstruct output masks through the skip connection (see the results in Table~\ref{tab:CT_MRI_results}).  
Again, it is seen from Fig.~\ref{fig:Box_CT_MRI_qual} that the DSC, VOE, and Sn from the EDNet are sparse for both the CT and MRI scans, with significantly fewer median metrics, which demonstrates that EDNet produces scattered results for each of the testing cases. On the other hand, the upper- and lower-whisker for all three metrics are closer for VGG-FCN8s and VGG-UNet than the EDNet (see in Fig.~\ref{fig:Box_CT_MRI_qual}), which shows better-robustness of them comparing the EDNet. 

Furthermore, the qualitative results in Fig.~\ref{fig:quali_CT_MRI} depict that both the VGG-UNet and VGG-FCN8s provide better-segmented substructures of both the modalities (CT and MRI) than the EDNet. In some examples of the EDNet, none of the target substructures are segmented, which provide VOE as $100.0\,\%$ (see upper-whisker in Fig.~\ref{fig:Box_CT_MRI_qual} (b)). The encoders in all three networks have 13-convolutional and 5-pooling layers, where the in-depth features obtained from them have lost spatial location information due to pooling in the encoders.
Hence, the output masks from EDNet have less local information, which is solved in the VGG-UNet and VGG-FCN8s due to the concatenation of local information through the skip connection, providing an alternative path for the gradient with backpropagation. It is experimentally validated in our investigations that the additional skipping paths are beneficial for improving the segmentation results of different heart substructures.

Again, it is also remarkable that VGG-FCN8s even exceeds the VGG-UNet (see the corresponding results in Table~\ref{tab:CT_MRI_results}, Fig.~\ref{fig:Box_CT_MRI_qual}, and Fig.~\ref{fig:quali_CT_MRI}) by the margins of $2.1\,\%$ and $0.6\,\%$ respectively for CT and MRI scans in terms of mDSC. The former network further outperforms the latter network concerning the other two metrics, such as mVOE and mSn, with significant margins for the CT scans.
The concatenation of low-level features from the encoder's antecedent layers with the equivalent decoder scale in VGG-UNet is the possible reason for failing VGG-UNet than VGG-FCN8s \cite{zhou2019unet++,hasan2020drnet}. 
Unessentially, an aggregation of the corresponding scale feature maps from the beginning layer of the encoder, is observed as a weakness of the UNet, as it imposes an undesirable merging procedure, forcing aggregation barely at the corresponding scale feature maps of the encoder and decoder, which was similarly experimentally validated in \cite{zhou2019unet++} and our previous article for other medical imaging modality \cite{hasan2020drnet}. 
On the other hand, in VGG-FCN8s, we fuse the feature maps from the encoder's different coarseness starting from the third block of the encoder, making it a winner of three implemented CNN-based WHS approaches.      

\subsubsection{Results for MALP-based methods}
\label{Results_for_MALP_based_methods}
The label propagated segmentation results utilizing different deformed atlas images are quantitatively and qualitatively manifested in Table~\ref{tab:CT_MRI_results}, Fig.~\ref{fig:Box_CT_MRI_qual}, and Fig.~\ref{fig:quali_CT_MRI}. 
The median and MVF schemes of level propagation, as described in subsection~\ref{Segmentation_Strategies}, generate the heart segmentation results with the mVOEs of $29.1\,\%$ and $25.1\,\%$, respectively, for CT scans and $55.6\,\%$ and $44.5\,\%$, respectively, for MRI scans. 
Table~\ref{tab:CT_MRI_results} demonstrates that the MVF scheme outperforms the median method of MALP by the margins of $2.7\,\%$ and $10.1\,\%$ respectively for CT and MRI modalities for mDSC. The former MALP method also outperforms the latter MALP method with significant margins concerning the other two metrics (mVOE and mSn).  
The Box and Whisker visualization of all three metrics in Fig.~\ref{fig:Box_CT_MRI_qual} for both the methods demonstrate the superiority of the MVF scheme than the median strategy for both the imaging modalities. 
Fig.~\ref{fig:quali_CT_MRI} (top) and Fig.~\ref{fig:quali_CT_MRI} (bottom) exhibit the qualitative results respectively for CT and MRI scans for both the median (MALP$_1$) and MVF (MALP$_2$) schemes. 
Those results qualitatively confirm that the segmented substructures from MALP$_1$ suffer from the outliers (see in fifth and sixth columns of Fig.~\ref{fig:quali_CT_MRI}), where most of the organs are erroneously labeled with other organs.  
All the experimental results reveal that the MVF scheme has better dealt with the outliers as it counts the majority of the labels from the voting candidates, whereas the median method estimates the median values of the counter, which may not be matched by the majority voters.  

\subsubsection{Results for probabilistic atlas-based methods}
\label{Results_for_probabilistic_atlas_based_methods}
The WHS results of the probabilistic atlas are exhibited in Table~\ref{tab:CT_MRI_results}, Fig.~\ref{fig:Box_CT_MRI_qual}, and Fig.~\ref{fig:quali_CT_MRI}, where we employ our three methods, such as PAS, EM, and PAS+EM (see details in subsection~\ref{Segmentation_Strategies}). 
The PAS, EM, and PAS+EM schemes of probabilistic atlas provide the heart segmentation results with the mVOEs of $14.5\,\%$, $87.8\,\%$, and $33.1\,\%$, respectively, for CT scans and $29.1\,\%$, $91.7\,\%$, and $52.7\,\%$, respectively, for MRI scans. 
Table~\ref{tab:CT_MRI_results} exhibit that the PAS scheme exceeds the other two methods, such as EM and PAS+EM, by the margins of $70.5\,\%$ and $12.8\,\%$ for CT scans concerning the mDSC, respectively, whereas it also beats them by the margins of $66.7\,\%$ and $20.0\,\%$ for MRI scans, respectively, in terms of the mDSC. Similarly, the EM and EM+PAS are also defeated by the proposed PAS method for mVOE and mSn for both the chest imaging modalities with considerable margins (see in Table~\ref{tab:CT_MRI_results}). 
The spreads and centers of the DSC, VOE, and Sn metrics for all the 20 test images (either CT or MRI scans), as exhibited in Fig.~\ref{fig:Box_CT_MRI_qual}, also reveal the supremacy of the proposed PAS method over the other two methods (EM and EM+PAS).  

The proposed PAS technique estimates the voxel class and assigns a target voxel label from the posterior distribution of the Bayesian framework, where the posterior distribution is determined from the likelihood tissue model and prior anatomical knowledge by a  MAP rule. Such an estimation of posterior distribution quantifies its expected probability value and the uncertainty associated with it, which probably makes the PAS algorithm a winner than the EM algorithm.  
As our EM algorithm is a single variate, it has less representation of the heart substructure features. The EM algorithm is prone to converging to local minima when dealing with a single variate dataset, which provides an overfitted model for the heart segmentation.

However, incorporating the PAS algorithm with an EM algorithm extends the WHS outcomes by the margins of $57.7\,\%$ and $46.7\,\%$ in terms of mDSC respectively for CT and MRI modalities. Still, the PAS technique is a defeating method.     
Again, Fig.~\ref{fig:quali_CT_MRI} (top) and Fig.~\ref{fig:quali_CT_MRI} (bottom) show the qualitative WHS outcomes for the CT and MRI scans for the PAS, EM, and PAS+EM designs, respectively. Those results show that the segmented substructures from EM and PAS+EM yield the outliers substructures (see in eight and ninth columns of Fig.~\ref{fig:quali_CT_MRI}), where the EM results suffer severely by the outliers. The embodiment of PAS with the EM has significantly reduced the outlier effect from the WHS results of the EM technique, as qualitatively depicted in Fig.~\ref{fig:quali_CT_MRI}.     

\subsection{Results comparison}
\label{Result_comparisons}
Comparing all the results as mentioned earlier, the proposed PAS is the best performing method for the WHS for both the utilized dataset modalities, such as CT and MRI scans, which are manifested and visualized in Table~\ref{tab:CT_MRI_results}, Fig.~\ref{fig:Box_CT_MRI_qual} and Fig.~\ref{fig:quali_CT_MRI}. 
The proposed and designed methods are classified into three categories: CNN-based, MALP-based, and probabilistic atlas-based, where the former approach works on the 2D slice of the 3D CT and MRI modalities, while the former two methods work on the whole 3D CT and MRI volumes. The experimentation reveals that the 3D image-based approach conquers the 2D image-based techniques.  
Our experimental WHS results also demonstrate that the 3D atlas-based approaches beat CNN-based 2D methods.  This result's probable reasons are that we take only axial slices from the 3D CT and MRI scans to train and evaluate the network. Although the 2D CNNs have much lighter computation and higher inference speed, it neglects the information between adjacent slices, which hinders the improvement of segmentation accuracy. Hence, the 3D CNNs can be a powerful model for learning representations for volumetric data and perceiving the volumetric spatial information, which will be analyzed for the same WHS task in the future.

\begin{figure}[!ht]
  \centering
  \subfloat{\includegraphics[width=8cm, height=5cm]{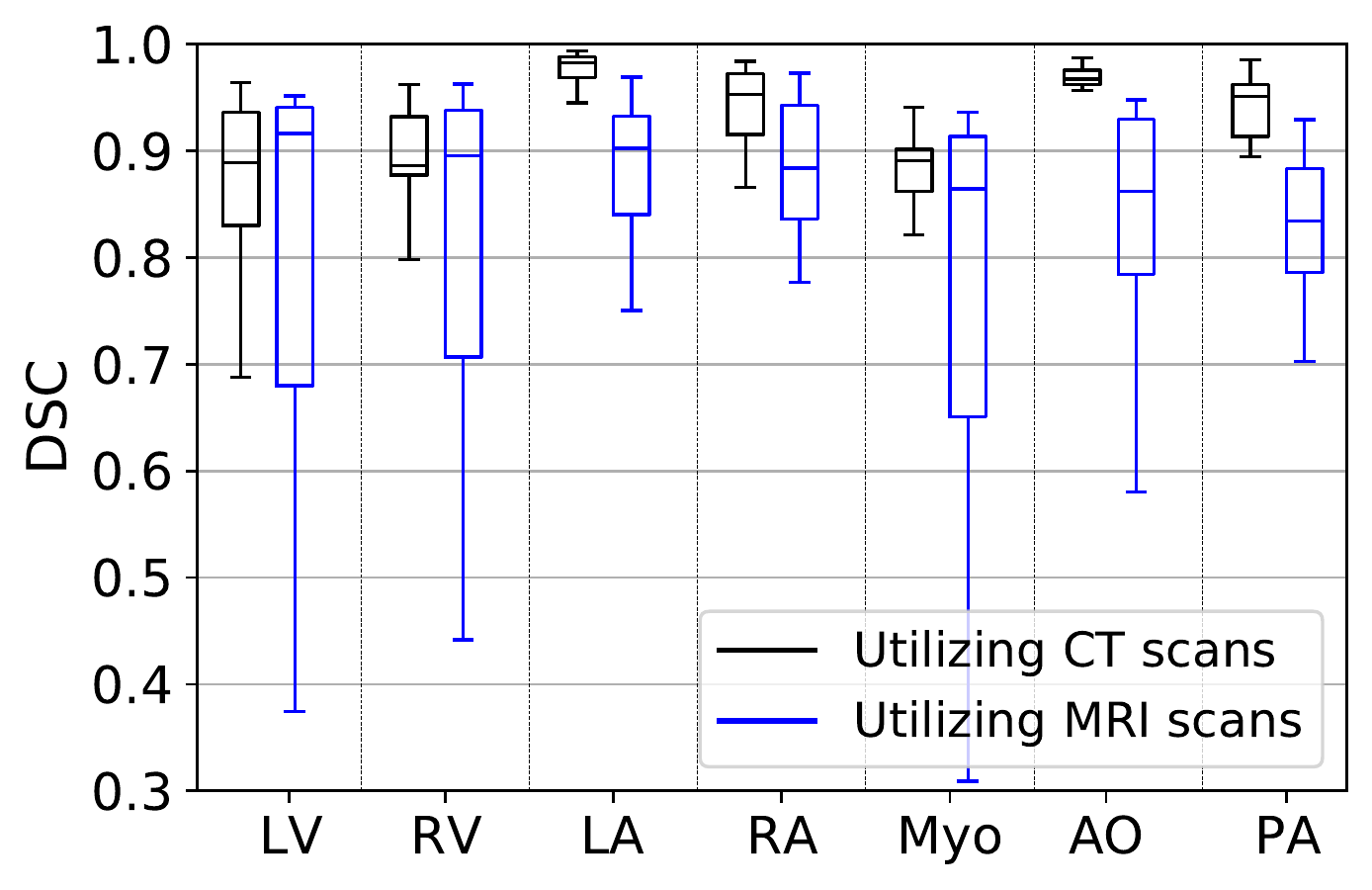}}
  \caption{The Box and Whisker visualization of DSC each substructure segmentation of WHS utilizing CT (black) and MRI (blue) scans. Our best performing segmentation method (PAS) for CT and MRI scans is employed to obtain this visualization.}
  \label{fig:Box_CT_MRI}
\end{figure}

Fig.~\ref{fig:Box_CT_MRI} reflects the Box and Whisker presentation of the mDSC of our proposed PAS algorithm on both the CT and MRI modalities for seven different substructures of the heart (see subsection~\ref{Problem_presentation}).  
The CT scans produce the best WHS results than the MRI scans for all the substructures in terms of mean and median values of the mDSC. The substructure's segmentation results utilizing the MRI images also have a higher interquartile range than the CT images (see in Fig.~\ref{fig:Box_CT_MRI}), which indicates less robustness of the MRI images for the WHS. 
WHS's poor results utilizing the MRI images over CT images are also reported in previously published articles \cite{payer2017multi, mortazi2017multi, liao2020mmtlnet, zhuang2019evaluation}.
The segmentation results, especially for LA, RA, Myo, AO, and PA, utilizing the CT images, are praiseworthy as they have a high median value and significantly less interquartile range than MRI images. 
Although MRI images' performance is less than CT, merging them can further improve the WHS results with the multi-modal input data, as it will have better feature presentations of different heart substructures \cite{liao2020mmtlnet}. Therefore, the future working direction can concentrate on the multi-modal utilization of the MM-WHS-2017 dataset applying this article's proposed techniques.     

Table~\ref{tab:resultcompare} shows the WHS results from our best performing method and other state-of-the-art methods on the same dataset utilizing both the CT and MRI scans.   
\begin{table*}[!ht]
\caption{Comparative results of our best performing method for WHS and state-of-the-art methods for the same task on both the MRI and CT scans and same dataset. Best and second-best metrics are underlined and double-underlined, respectively.}
\centering
\begin{tabular}{lccccc}
\hline
\rowcolor[HTML]{C0C0C0} 
\cellcolor[HTML]{C0C0C0}                                    & \cellcolor[HTML]{C0C0C0}                       & \cellcolor[HTML]{C0C0C0}                           & \multicolumn{3}{c}{\cellcolor[HTML]{C0C0C0}Metrics} \\ \cline{4-6} 
\rowcolor[HTML]{C0C0C0} 
\multirow{-2}{*}{\cellcolor[HTML]{C0C0C0}Different methods} & \multirow{-2}{*}{\cellcolor[HTML]{C0C0C0}Year} & \multirow{-2}{*}{\cellcolor[HTML]{C0C0C0}Modalities} & mVOE             & mSn            & mDSC            \\ \hline

Multi-scale patch and multi-modality atlas-based WHS \cite{zhuang2016multi} &                                              $2016$  &     CT+MRI                                               &           $0.183$       &   $-$             &     $0.899$             \\

Multi-atlas registration-based approach \cite{heinrich2017mri} &       $2017$                                         &     MRI                                         &          $0.246$          &     $-$         &    $0.860$            \\

 Multi-planar CNN   \cite{mortazi2017multi} &                                              $2017$  &     MRI                                               &           $0.259$       &  $0.831$               &     $0.851$             \\ 

 Multi-planar CNN   \cite{mortazi2017multi} &                                               $2017$  &       CT                                             &    $0.187$              &    $0.866$            &    $0.897$  \\

Multi-label FCN \cite{payer2017multi} &     $2017$                                           &     MRI                                            &          $0.230$        &    $-$          &     $0.870$            \\ 

Multi-label FCN \cite{payer2017multi} &       $2017$                                         &     CT                                            &         $0.168$         &      $-$        &    $0.908$             \\

CFUN: faster R-CNN + 3D UNet  \cite{xu2018cfun} &       $2018$                                         &     CT                                            &         $0.247$         &    $-$        &    $0.859$             \\

HFANet: deep heterogeneous feature aggregation network \cite{zheng2019hfa} &  $2019$                                              &     CT                                            &     $0.167$             &     $-$        &    $0.909$             \\

Two-stage UNet with adaptive threshold window \cite{liu2019automatic} &     $2019$      &     CT+MRI       &            $0.343$      &      $-$        &     $0.793$            \\

Multi-depth fusion of 3D UNet combining local and global features \cite{ye2019multi} &       $2019$                                         &     CT                                         &          $0.170$          &    $-$         &    $0.907$           \\

Averaging of 10 different algorithm \cite{zhuang2019evaluation} &     $2019$                                           &     MRI                                            &    $0.299$              &    $-$          &    $0.824$         \\

Averaging of 10 different algorithm \cite{zhuang2019evaluation} &       $2019$                                         &     CT                                         &          $0.227$          &    $-$        &    $0.872$             \\

Dual-Teacher: integrating intra-domain and
inter-domain teachers \cite{li2020dual} &                                   $2020$             &     CT+MRI                                               &           $0.245$       &    $-$        &  $0.860$               \\

MMTLNet: multi-modality transfer learning with adversarial training \cite{liao2020mmtlnet} &     $2020$                                           &     MRI                                            &    $0.198$              &     $-$         &     $0.890$           \\ 
 
MMTLNet: multi-modality transfer learning with adversarial training \cite{liao2020mmtlnet} &       $2020$                                         &     CT                                         &          \doubleunderline{$0.158$}          &     $-$          &    \doubleunderline{$0.914$}             \\

MvMM-RegNet: multivariate mixture model with registration framework \cite{luo2020mvmm} &     $2020$                                           &     MRI                                            &    $0.229$               &   $-$           &    $0.871$          \\ 
 
3D UNet incorporating of the principal component analysis for augmentation \cite{habijan2020neural} &       $2020$                                         &     CT                                         &          $0.211$          &   $-$       &   $0.882$             \\

\textbf{Our proposed probabilistic atlas-based WHS}  &     2020                                            &  MRI                                                  &           $0.291$       &        $0.825$        &     $0.819$            \\

\textbf{Our proposed probabilistic atlas-based WHS} &      2020                                         &      CT                                              &   \underline{$0.145$}               &        \underline{$0.919$}        &     \underline{$0.921$}            \\ \hline

\end{tabular}          
\label{tab:resultcompare}
\end{table*}
The comparative results, as in Table~\ref{tab:resultcompare}, demonstrate that the proposed PAS method with the CT scans outperforms the recent state-of-the-art approach in \cite{liao2020mmtlnet} by the margins of $1.3\,\%$ and $0.7\,\%$ concerning the mVOE and mDSC, respectively. 
The authors in the second-best method (MMTLNet) \cite{liao2020mmtlnet} transferred the MRI image information from the source domain to the target CT domain through the adversarial training without considering their spatial alignment, which could be the viable reasons for realizing the WHS multi-modality approach.  
The other two well-performing methods (see in Table~\ref{tab:resultcompare}), such as Multi-label FCN \cite{payer2017multi} and Multi-depth fusion of 3D UNet combining local and global features \cite{ye2019multi}, also carry possible drawbacks for being defeated by our proposed approach. 
The former strategy applied two stages, wherein the first stage, the VOIs are selected for the essential second stage. The errors in VOI selection could lead to erroneous heart segmentation results.
The latter approach employed the fusion from the different depths of the CNN network, where they fused the input block to the along with other depths to generate the final WHS VOIs. The concatenation of the input block with the same scale outermost decoder's block probably hamper the precise output, as it is also proven in our VGG-UNet and VGG-FCN8s experiments (see in Table~\ref{tab:CT_MRI_results}), and the articles in \cite{zhou2019unet++,hasan2020drnet}.

\section{Conclusions and Future Work}
\label{Conclusions}
Accurate in the whole heart's segmentation is crucial in developing clinical applications, although it is very challenging due to diverse artifacts, image variability, and patient-specific properties. 
This article has introduced and explored a robust and accurate pipeline for the WHS utilizing two different heart imaging modalities, such as CT and MRI scans. 
It is experimentally validated that incorporating the prior anatomical knowledge represented by probabilistic atlas into the Bayes inference to delineate seven different heart substructures has better-segmented results while utilizing the CT scans and employing our multi-resolution non-rigid registration pipeline.  
As the 2D CNNs fail to provide satisfactory WHS results, future research will focus our research direction on training and evaluating networks in multiple directions (coronal, sagittal, and axial) to combine all the directional WHS results. We will also design an end-to-end 3D segmentation network for comprehensive ablation studies. The recommended pipeline will be applied to other domains for volumetric medical image segmentation to verify its versatility and generability.

\section*{Conflict of Interest}
The authors have not any conflicts to disclose this research.

\bibliographystyle{IEEEtran} 

\bibliography{refs}

\begin{IEEEbiography}[{\includegraphics[width=1.05in,height=1.2in]{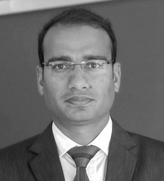}}]{Tarun Kanti Ghosh} was born in Jashore, Khulna, Bangladesh, in 1983 and earned the B.S. and M.S. degrees in biotechnology and genetic engineering from the Khulna University of Khulna, Bangladesh, in 2011 and another M.S. degree in biomedical engineering from
Khulna University of Engineering \& Technology (KUET), Khulna, Bangladesh, in 2016. From 2006 to 2008, he was a Research Fellow with the Animal Cell Culture and Molecular Biology Laboratory. He is currently a Ph.D. student in biomedical engineering at KUET.  He is the author of three international journal and conference articles. His research interests include biomedical signal and image processing, bioelectricity, bioinformatics, biomedical systems modeling and simulation, computational intelligence in biomedical engineering.
\end{IEEEbiography}

\begin{IEEEbiography}[{\includegraphics[width=1.05in,height=1.2in]{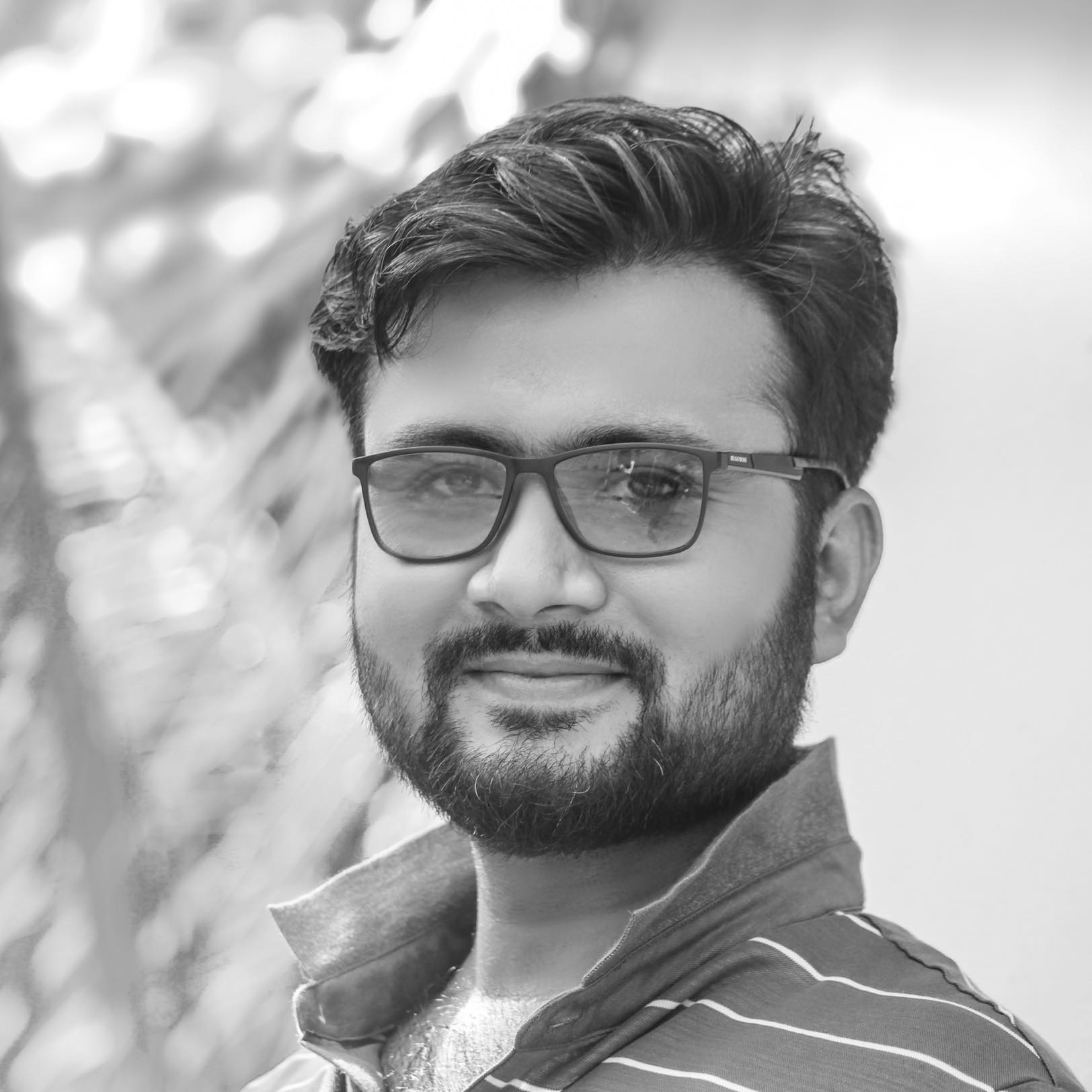}}]{Md. Kamrul Hasan} was born in Tangail, Bangladesh, in 1992 and received B. Sc. and M. Sc. engineering degrees in Electrical and Electronic Engineering (EEE) from Khulna University of Engineering \& Technology (KUET) in 2014 and 2017, respectively. He received another M. Sc. in Medical Imaging and Application (MAIA) from France (University of Burgundy), Italy (University of Cassino and Southern Lazio), and Spain (University of Girona) as an Erasmus scholar in 2019. Currently, Mr. Hasan is serving as an Assistant Professor at KUET in the EEE department. He analyzed different medical image modalities and machine learning during the MAIA study to build a generic computer-aided diagnosis system. His research interest includes medical image and data analysis, machine learning, deep convolutional neural network, medical image reconstruction, and surgical robotics in minimally invasive surgery. Mr. Hasan is currently a supervisor of several undergraduate students on the classification, segmentation, and registration of medical images with different modalities. He has already published many research articles on medical image and signal processing in different international journals and conferences.
\end{IEEEbiography}

\begin{IEEEbiography}[{\includegraphics[width=1.05in,height=1.2in]{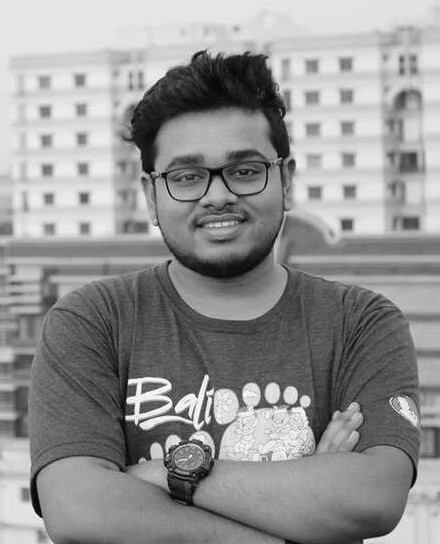}}]{Shidhartho Roy} received the B.Sc. degree in electrical and electronic engineering from the Khulna University of Engineering \& Technology (KUET), in 2020. His undergraduate thesis was on “EEG based Brain-Computer Interaction with different stimuli using machine learning.” He is currently working on Bio-signal processing, Medical Imaging, 3D Reconstruction of Medical Images, Renewables, and Computer-Assisted Interventions. His research interests include artificial intelligence, brain-computer interface, and deep learning. Mr. Roy awarded seven national awards so far for his work.  His previous works were presented at TENSYMP 2020, ICAEE 2019, ICIECE 2019, EICT 2019, and Artificial Intelligence in Medicine (AIIM, Elsevier).
\end{IEEEbiography}

\begin{IEEEbiography}[{\includegraphics[width=1.05in,height=1.2in]{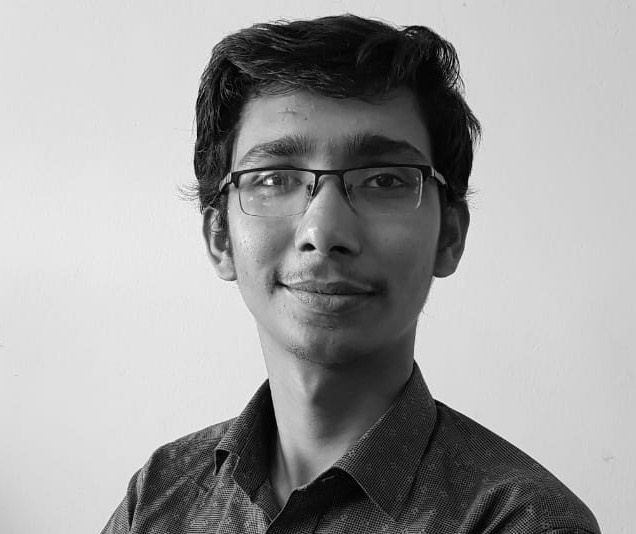}}]{Md. Ashraful Alam} is studying Electrical and Electronic Engineering (EEE) at Khulna University of Engineering \& Technology (KUET). Currently, he is working on  Medical Imaging. His interests include medical image and data processing, computer vision, and deep learning. Mr. Alam awarded two national awards so far for his work in the idea development project. His previous works were presented at IEEE Access and Artificial Intelligence in Medicine (AIIM, Elsevier).
\end{IEEEbiography}

\begin{IEEEbiography}[{\includegraphics[width=1.05in,height=1.2in,clip,keepaspectratio]{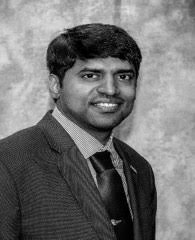}}]{Eklas Hossain} (M’09, SM’17) received his Ph. D. from the College of Engineering and Applied Science at University of Wisconsin Milwaukee (UWM). He received his MS in Mechatronics and Robotics Engineering from International Islamic University of Malaysia, Malaysia, in 2010 and BS in Electrical and Electronic Engineering from Khulna University of Engineering \& Technology, Bangladesh, in 2006. Dr. Hossain has been working in distributed power systems and renewable energy integration for the last ten years, and he has published several research papers and posters in this field. He is now involved with several research projects on renewable energy and grid-tied microgrid system at Oregon Tech, as an Assistant Professor in the Department of Electrical Engineering and Renewable Energy since 2015. 
He is a senior member of the Association of Energy Engineers (AEE). He is currently serving as an Associate Editor of IEEE Access. He is working as an Associate Researcher at the Oregon Renewable Energy Center (OREC). He is a registered Professional Engineer (PE) in the state of Oregon, USA. He is also a Certified Energy Manager (CEM) and Renewable Energy Professional (REP). His research interests include modeling, analysis, design, and control of power electronic devices; energy storage systems; renewable energy sources; integration of distributed generation systems; microgrid and smart grid applications; robotics, and advanced control system. He is the winner of the Rising Faculty Scholar Award in 2019 from the Oregon Institute of Technology for his outstanding contribution to teaching. With his dedicated research team, Dr. Hossain is looking forward to exploring methods to make electric power systems more sustainable, cost-effective, and secure through extensive research and analysis on energy storage, microgrid system, and renewable energy sources.
\end{IEEEbiography}

\begin{IEEEbiography}[{\includegraphics[width=1.05in,height=1.2in]{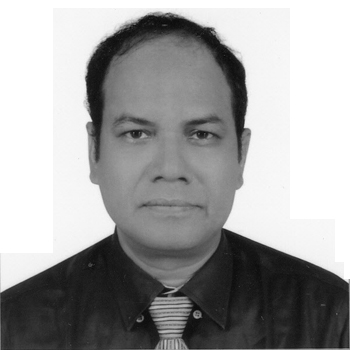}}]{Mohiuddin Ahmad} received his BS degree with Honors Grade in Electrical and Electronic Engineering (EEE) from Chittagong University of Engineering and Technology (CUET), Bangladesh, and his MS degree in Electronics and Information Science (major – Biomedical Engineering)
from Kyoto Institute of Technology of Japan in 1994 and 2001, respectively. He received his Ph.D. degree in Computer Science and Engineering (CSE) from Korea University, Republic of Korea, in
2008. From November 1994 to August 1995, he served as a part-time Lecturer in the Department of EEE at CUET, Bangladesh. From August 1995 to October 1998,
he served as a Lecturer in the Department of EEE at Khulna University of Engineering \& Technology (KUET), Bangladesh. In June 2001, he joined the same Department as an Assistant Professor. In May 2009, he joined the same Department as
an Associate Professor, and now he is a full professor. 
Moreover, Dr. Ahmad served as the Head of the Department of Biomedical Engineering (BME) from October 2009 to September 2012. Prof. Ahmad served as the Head of the Department of EEE from September 2012 to August 2014. From July 2014, Prof. Ahmad has been serving as the sub-project manager of the UGC, HEQEP, Sub-Project, CP\#3472, titled on Postgraduate Research in
BME. 
His research interests include biomedical signal and image processing for disease diagnosis, computer vision and pattern recognition, clinical engineering \& modern healthcare. Prof. Ahmad is
the life-fellow of IEB and a member of IEEE.
\end{IEEEbiography}

\EOD

\end{document}